%% file: ruchika_paper.tex
\documentclass[fleqn,usenatbib]{mnras}

\usepackage{newtxtext,newtxmath}

\usepackage[T1]{fontenc}
\usepackage{ae,aecompl}

\usepackage{graphicx}	
\usepackage{amsmath}	
\usepackage[utf8, latin1]{inputenc}
\usepackage{longtable}
\usepackage{threeparttable}
\usepackage{threeparttablex}
\usepackage{color}
\usepackage{hyperref}
\usepackage[dvipsnames]{xcolor}

\newcommand{\M}{\ensuremath{M_{\odot}}}
\newcommand{\Ll}{\ensuremath{L_{\odot}}}
\newcommand{\chandra}{\emph{Chandra}}
\newcommand{\xmm}{\emph{XMM-Newton}}

\title[Non-central radio galaxies in clusters]{The contribution of non-central radio galaxies to AGN feedback in rich galaxy clusters}

\author[Seth et al.]{
Ruchika Seth$^{1}$\thanks{Deceased},
Ewan O'Sullivan$^{2}\thanks{eosullivan@cfa.harvard.edu}$,
Biny Sebastian$^{3}$,
Somak Raychaudhury$^{1,4}$,
\newauthor Gerrit Schellenberger$^{2}$, 
Christopher P. Haines$^{5}$
\\
$^{1}$Inter-University Centre for Astronomy and Astrophysics, Savitribai Phule Pune University Campus, Ganeshkhind, Pune 411007, India\\
$^{2}$Center for Astrophysics $|$ Harvard \& Smithsonian , 60 Garden Street Cambridge, MA 02138, USA\\
$^{3}$Department of Physics and Astronomy, Purdue University, 525 Northwestern Avenue, West Lafayette, IN 47907, USA\\
$^{4}$School of Physics and Astronomy, University of Birmingham, Birmingham B15~2TT, UK\\
$^{5}$Instituto de Astronom\'{i}a y Ciencias Planetarias de Atacama, Universidad de Atacama, Copayapu 485, Copiap\'{o}, Chile\\
}

\date{Accepted 2022 April 11. Received 2022 April 11; in original form 2021 September 16}

\pubyear{2022}

\begin{document}
\label{firstpage}
\pagerange{\pageref{firstpage}--\pageref{lastpage}}
\maketitle

\begin{abstract}
We present a combined radio/X-ray study of six massive galaxy clusters, aimed at determining the potential for heating of the intra-cluster medium (ICM) by non-central radio galaxies. Since X-ray cavities associated with the radio lobes of non-central galaxies are generally not detectable, we use Giant Metrewave Radio Telescope 610~MHz observations to identify jet sources and estimate their size, and \chandra\ data to estimate the pressure of the surrounding ICM. In the radio, we detect 4.5\% of galaxies above the spectroscopic survey limit (M$^*_{K}$+2.0) of the Arizona cluster redshift survey (ACReS) which covers five of our six clusters. Approximately one tenth of these are extended radio sources. Using star formation (SF) rates determined from mid-infrared data, we estimate the expected contribution to radio luminosity from the stellar population of each galaxy, and find that most of the unresolved or poorly-resolved radio sources are likely star formation dominated. The relatively low frequency and good spatial resolution of our radio data allows us to trace star formation emission down to galaxies of stellar mass $\sim$10$^{9.5}$ \M{}. We estimate the enthalpy of the (AGN dominated) jet/lobe and tailed sources, and place limits on the energy available from unresolved radio jets. We find jet powers in the range $\sim$10$^{43}$-10$^{46}$~erg~s$^{-1}$, comparable to those of brightest cluster galaxies. Our results suggest that while cluster-central sources are the dominant factor balancing ICM cooling over the long term, non-central sources may have a significant impact, and that further investigation is possible and warranted.
\end{abstract}

\begin{keywords}
galaxies: clusters: general -- galaxies: active -- galaxies: jets -- galaxies: statistics -- radio continuum: galaxies -- galaxies: clusters: intracluster medium
\end{keywords}



\section{Introduction}

The fundamental characteristic of cool-core galaxy clusters is that the central cooling times of their hot, X-ray emitting intra-cluster medium (ICM) are substantially less than the Hubble time. The gas in the cores of such clusters cools, primarily via thermal bremsstrahlung radiation in the X-ray band, leading to a change in the local pressure gradient. This results in the ICM gas flowing in towards the cluster core. The subsequent compression leads to an enhanced density, which in turn results in a faster cooling (since the rate of thermal bremsstrahlung radiation is proportional to density squared) following the cooling-flow model \citep{Fabianetal84}. In principle, this process should lead to a build up of cool gas in the cluster core, fuelling star formation. Observationally, while some cluster cores do contain cool gas, the quantity is far less than simple models would predict, as are the  star-formation rates \citep[e.g.,][]{McDonaldetal18} found in brightest cluster galaxies (BCGs).

It is now widely accepted that heating by the active galactic nuclei (AGN) of cluster-central galaxies is the dominant process balancing this radiative cooling \cite[e.g.,][among others]{Raffertyetal06, Birzanetal08, OSullivanetal11b, McNamaraetal06}. Extensive observational evidence shows that the jets launched by the central AGN can impact the surrounding ICM, and that the activity of the nucleus is strongly correlated with ICM properties \citep[e.g.,][]{McNamaraNulsen12,Gittietal12,Fabian12,Eckertetal21}. The AGN of cluster-dominant galaxies are typically found to be  Fanaroff-Riley type~I sources \citep[FR-I,][]{FanaroffRiley74}, though cluster-dominant FR-IIs are not unknown \citep[e.g., Cygnus~A, 3C~220.1; ][]{Wilsonetal06,Sniosetal18,Liuetal20}. In both cases, the expansion of the radio jets into the ICM and the inflation of radio lobes is thought to inject energy into the ICM via \textit{PdV} heating, shocks, compression, and increased turbulence.

Numerous studies have shown depressions in X-ray surface brightness correlated with the radio lobes of BCGs \citep[e.g.,][]{Vrtileketal00,FinoguenovJones01,Formanetal05,McNamaraetal05,Dunnetal05,Dongetal10,Shinetal16}. The enthalpy of these cavities, consisting of the work done on the ICM by the lobes as they expand and the energy stored within them, can be of the order of 10$^{62}$~erg \citep[e.g.,][]{McNamaraNulsen07} sufficient to balance radiative losses and thus halt (or dramatically slow) cooling \citep{Birzanetal04,Raffertyetal06,Panagouliaetal14b}. Cavities are observed in $\sim$70 per cent of nearby cool core systems \citep{Dunnetal05}.

While the centres of galaxy clusters and groups are known to be an especially favourable location for AGN, with their central galaxies significantly more likely to host radio sources than other early-type galaxies of similar mass \citep[e.g.,][]{Bestetal07,LinMohr07,Smolcicetal11}, non-central cluster galaxies can also host radio jet sources \citep[e.g.,][]{Clarkeetal19,GendronMarsolaisetal20}. The presence of a radio AGN is strongly correlated with galaxy mass \citep{Bestetal05,Sabateretal19}, with the most massive ellipticals showing near-constant radio activity. Estimates of the typical power output of radio jets across the whole population of early-type galaxies suggests that they should be sufficient to balance radiative losses in their own halos, across the range of mass scales from individual galaxies to galaxy clusters \citep{Bestetal07,Hardcastleetal19}. Galaxy clusters often contain multiple early-type galaxies with masses greater than a few 10$^{10}$\M, and these have the potential to launch powerful radio jets which can extend beyond the body of the galaxy and interact with the surrounding ICM.

In this paper we present a pilot study, aimed at determining the contribution to ICM heating from non-central cluster radio galaxies. Direct detection of cavities associated with non-central galaxies in X-ray imaging is challenging. X-ray emissivity falls off as the square of the density, thus for sources at large cluster-centric radii the expected contrast between a cavity and the surrounding ICM emission can be quite low. For nearby clusters, where expected fluxes will be higher, non-central sources may fall outside the fields for which X-ray data with sufficient angular resolution to detect cavities are available. Cavities associated with non-central cluster galaxies are not unknown, e.g., in the Virgo cluster galaxies NGC 4477 \citep{Lietal18b} and NGC~4552 \citep{Machaceketal06}, but in these systems the cavities are relatively small (1-2~kpc across) and appear to be contained within the denser hot halos of their host galaxies. Cavities of sufficient size will of course be detectable in deep X-ray data even at large radius. The ancient cavity in the outskirts of the Ophiuchus cluster is one example; it is $\sim$430~kpc across, and even at a mean radius of $\sim$345~kpc from the cluster centre its inner boundary is clearly visible in a relatively modest 37~ks \xmm\ observation \citep{Giacintuccietal20}. However, this is an exceptional source and the cavities of most non-central galaxies would be expected to be considerably smaller. Using volumes estimated from non-central radio sources considered in two of the clusters used in this study (ZwCl~3146 and RXJ2129), we estimate the significance with which spherical cavities might be detected in \chandra\ data, and find that observations 3-10 times deeper than those currently available, with total exposures of several hundred kiloseconds, would be required for 3$\sigma$ detections. In practice, given non-ideal cavity morphologies and uncertainties about the radio source positions within their clusters, megasecond exposures would likely be needed. It is impractical to build up a sample of clusters observed in such depth with current observatories.

However, radio observations provide an effective alternative mechanism to identify AGN and determine their extent and morphology. We therefore make use of a combination of Giant Metrewave Radio Telescope (GMRT) and \chandra\ X-ray observations to locate AGN candidates and measure ICM properties. Cluster membership and galaxy properties are established using deep optical spectroscopic and multi-wavelength photometric catalogues.

This paper has been organised in the following manner: \S~\ref{data} describes the cluster sample selection and the spectroscopic catalogue, \S~\ref{radio data analysis} and \S~\ref{x-ray data analysis} describe the various tools and techniques used for the radio and X-ray data reduction, \S~\ref{results} outlines the main results and discusses their implications. \S~\ref{conclusions} presents the major conclusions of the paper.
Throughout the paper we adopt the standard $\Lambda$CDM Cosmology, assuming a flat universe with $\Omega_\Lambda\! =\! 0.7$, $\Omega_M\! =\! 0.3$ and $H_0 = 70 $ km~s$^{-1}$ Mpc$^{-1}$.

\section{Cluster sample}
\label{data}
Our sample consists of six cool-core clusters selected to be in both the Local Cluster Substructure Survey \citep[LoCuSS,][]{OkabeSmith16,Hainesetal13} and the \citet{Birzanetal04} sample of cool core clusters. Our systems were chosen to have at least 100 redshifts available, and to have  $>$4~hr GMRT 610~MHz observations performed in cycle~18 or later. LoCuSS is a sample of more than 100 X-ray-luminous clusters selected from the \textit{ROSAT} All-Sky Survey (RASS). It is a multi-wavelength survey primarily focused on calibrating the mass-observable scaling relations obtained from weak-lensing, X-ray or Sunyaev-Zel'dovich analyses \citep[e.g.][]{Mulroyetal19}. A particularly rich multi-wavelength dataset is available \citep{Hainesetal13}, which includes Subaru/Suprime-Cam optical imaging \citep{Okabeetal10}, Herschel/PACS+SPIRE 100-500 $\mu$m maps, Spitzer/MIPS 24 $\mu$m maps, near-infrared (NIR; J and K bands) imaging and \chandra~X-ray data. For all clusters in our sample the optical/IR data probe distances to $\sim$ 1.5-2 virial radii corresponding to a field of view $25'\times25'$. GMRT and \chandra\ observations of the six clusters are present in the respective archives. The presence of cool cores was confirmed from the Archive of \chandra\ Cluster Entropy Profile Tables (ACCEPT) database \citep{Cavagnoloetal09}.

Global cluster parameters, such as ICM temperature and the volume and power of cavities associated with the BCG, were obtained from \cite{Raffertyetal06} and \cite{Cavagnoloetal09}. Cluster size (R$_{200}$) and some other optical parameters were taken from the WHL catalogue \citep{Wenetal12}. The cluster properties and parameters are listed in Table.~\ref{tab:cluster_table}.

\input{tables/table1.tex}

The clusters in our sample cover a range of morphologies and dynamical states from fully relaxed clusters to merging systems close to core passage. None of the clusters shows any signs of core destruction due to shocks and turbulence in the ICM; all possess cool cores.

None of the Central BCGs in our sample hosts a prominent FR-I or FR-II radio galaxy. Two clusters in the sample (Abell\,1758 and Abell\,1914) are merger systems. The remainder, Abell\,383, RXJ\,1720, RXJ\,2129 and ZwCl\,3146, have relatively relaxed morphologies \citep{Owersetal11,Mantzetal14,Kaleetal15b,Savinietal19,ZuHoneSims19}. The two merging clusters host radio haloes \citep[e.g.][]{Schellenbergeretal19,Mandaletal19} and at least three of the four relaxed systems host radio mini-haloes \citep[RXJ~1720, RXJ~2129, ZwCl~3146,][]{Giacintuccietal14b,Kaleetal15b,Giacintuccietal19}. The merging cluster Abell~1914 also contains a radio phoenix \citep{Mandaletal19}.

\subsection{Spectroscopic cluster members}
\label{Optical_members}

Extensive spectroscopic data from the Hectospec multi-object spectrograph on the 6.5m MMT telescope are available for the galaxy populations of all six clusters, in five cases from the Arizona cluster redshift survey \citep[ACReS,][]{Hainesetal13}, and in the sixth, ZwCl~3146, from the Hectospec Cluster Survey  \citep[HeCS,][]{Rinesetal13}.

ACReS targeted likely cluster members brighter than $M^*_{K}+2.0$. Overall, the completeness of the spectroscopic survey is about 87\% for cluster galaxies brighter than $M^*_{K}+2.0$ and lying within $r_{200}$. The targets detected at 24~$\mu$m were prioritised to ensure a 100\% completeness for cluster members with $f_{24} > 0.4$ mJy. Each of the five clusters was observed in 4-5 independent configurations to minimise the overall effects of incompleteness and fibre collisions. At the same time, the previous targeted surveys from the literature provide additional redshifts to fill in any residual incompleteness remaining in the densest regions of the cluster cores. All six clusters are within the Sloan Digital Sky Survey (SDSS) footprint, providing additional redshifts and $ugriz$ photometry.

\subsection{Star formation rates}
\label{sec:SFR}

Star-formation is the dominant source of radio emission in lower stellar mass  \citep[$M_{*}{<}10^{11}{\rm M}_{\odot}$;][]{Bestetal05} and low to moderate radio luminosity \citep[$L_{1.4\,{\rm GHz}}{<}10^{23}$W\,Hz$^{-1}$;][]{Yunetal01} galaxies. We therefore need to account for its contribution in our clusters. Derived properties such as star formation rate (SFR) and stellar mass are available only for the five clusters present in the ACReS sample. To determine the star formation rates in these galaxies, observations of each cluster were performed at 24 $\mu$m with MIPS onboard the Spitzer Space Telescope \citep{Riekeetal04} across a $25'\times25'$ field-of-view, achieving 90\% completeness limits of 0.4\,mJy. A complementary view at 100-500\,$\mu$m of each cluster field was provided by Herschel PACS/SPIRE imaging reaching sensitivities comparable to the obscured star formation rates.  Total infrared luminosities inferred from the two telescopes enabled a distinction between the AGN at infrared wavelengths from star-formation. 

The luminosity-dependent SED templates of \citet{Riekeetal09}, matched to the observed 24 $\mu$m flux, gives the rest-frame 24-$\mu$m luminosity and the total infrared luminosity.  The SFRs are then inferred from this 24~$\mu$m rest-frame luminosity by using the conversion of \citet{Riekeetal09}, given by 

\begin{equation}
\frac{\rm SFR}{\M{} yr^{-1}} = 7.8\times 10^{-10} \frac{L_{24~\mu m}}{\Ll{}}
\end{equation} 

\noindent which is valid for both Kroupa or Chabrier initial mass functions (IMFs). The 24~$\mu$m data were sensitive down to SFRs of 1--2 \M{} yr$^{-1}$ among cluster galaxies.  

\subsection{Stellar Mass}
Stellar masses are derived from a combination of colour information from optical SDSS magnitudes \citep{Yorketal00} and $K$-band magnitude from LoCuSS. The mass-to-light ratio is given by log(M/L$_{K}$) = a$_{K}$ + (b$_{K}$ $\times$ (g-i)) where a$_{K}$ and b$_{K}$ are coefficients with a$_{K}$=$-0.211$ and b$_{K}$=$0.137$, $g$ and $i$ are SDSS magnitudes, and the M/L$_{K}$ is in solar units \citep{Belletal03}. The assumed `diet' Salpeter IMF was converted to a Chabrier IMF by subtracting 0.1 from the zero point (a$_{K}$). $K$-band light traces the underlying stellar mass, and so the colour information becomes less relevant when using the relationship of \citet{Belletal03}. The mass of the redder galaxies is over-estimated without the NIR data.  We estimated the mass-to-light ratios to be around 0.69--0.94 corresponding to the 10-90 per cent quantiles. The resulting stellar masses for the cluster members ranged from $\sim$ 10$^{10}$-10$^{12}$ \M{}.

\begin{figure*}
    \centering
    \includegraphics[width=\textwidth]{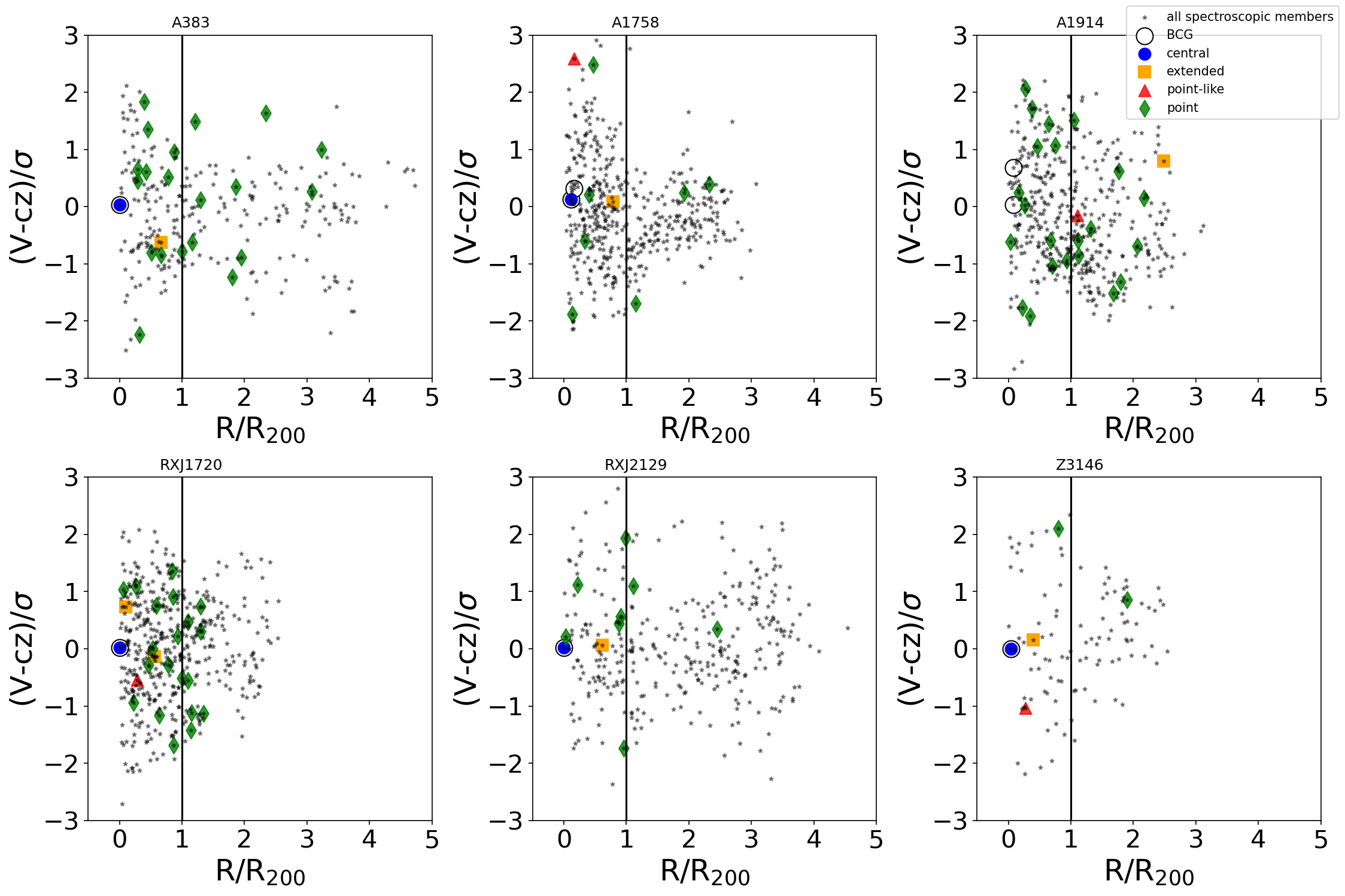}
    \caption{Distribution of all optically-confirmed cluster members in our six clusters in the projected radial distance and velocity space. Top panel (left to right): Abell\,383, Abell\,1758, Abell\,1914. Bottom panel (left to right): RXJ\,1720, RXJ\,2129 and ZwCl\,3146. BCGs are represented by black open circles. Radio-loud BCGs are coloured in blue. Note that Abell\,1914 is the only cluster in our sample which does not have any central radio source. Galaxies detected in radio are marked, with orange squares representing extended radio sources, green diamonds point sources, and red triangles point-like radio sources.}
    \label{fig:v_sigma_r_r200}
\end{figure*}

\input{tables/table3.tex}


\section{GMRT Data Analysis}
\label{radio data analysis}

\subsection{Calibration and Imaging}

Details of the GMRT observations used are given in Table~\ref{table:GMRTtable}. We used the Astronomical Image Processing System \citep[\textsc{aips},][]{Greisen90} and the \textsc{flagcal} package \cite{PrasadChengalur12} for flagging of the data contaminated by radio frequency interference (RFI). After the initial editing, standard \textsc{aips} tasks were used to carry out the basic flux, phase, and bandpass calibration. We used the \textsc{aips} task \textsc{split} to average channels and split out the target source after applying the basic calibrations. Three rounds of imaging and phase-only self-calibration were carried out before doing the final amplitude and phase self-calibration. While imaging, we used the faceting algorithm to take care of the w-term errors. Finally, we merged the various facets on a single plane using the \textsc{aips} task \textsc{flatn} and corrected for the primary beam using the \textsc{aips} task \textsc{pbcor}.

\subsection{Radio Sources in the GMRT images}
\label{subsection:pybdsf}
To identify sources in the radio images, we employed the Python Blob Detector and Source Finder ({\sc PyBDSF}) \footnote{For further details see~\url{http://www.astron.nl/citt/pybdsf/}. The code is available at ~\url{https://github.com/lofar-astron/PyBDSF}}. \textsc{PyBDSF} takes the radio image as its input and uses a wavelet-based algorithm to detect all sources with significant radio emission, above a threshold set by the user. The radio sources detected in the image were further decomposed into smaller components. These multiple sources were then fitted using multi-scale Gaussians to obtain a residual image with a mean of zero and a low variance. 

We employed a minimum island threshold of 6$\sigma$ for all images, where an island is defined in terms of a boundary enclosing sources of radio emission, whose integrated flux density is above the chosen threshold of the mean map of the image. Each island is composed of one or more discrete radio sources. To get an estimate of the  contribution from each of these sources to the observed flux, multiple Gaussians with varying parameters were fitted to these sources. The residual image is obtained after fitting and subtracting the required flux density of the source and determines the quality of the fit.

Based on the type and number of Gaussians employed to extract the flux information, each island can be classified into one of three categories; `S', `C' and `M'. Category `S' sources are described by a  single Gaussian model and occupy only a single island. They are mostly simple point sources. Category `C' sources are extended sources requiring multiple Gaussian components which may extend over multiple islands. Examples include a head-tail galaxy whose head and tail occupy two separate islands, or a double-lobed source whose lobes and core occupy separate islands. Category `M' sources are extended sources on a single island which require multiple Gaussian components to accurately describe their shape. An example might be a radio core with a small single-sided jet.

We note that the GMRT data for A1758 and ZwCl\,3146 have rms noise limits that are \textasciitilde{}7 times higher than those for A383, A1914 and RXJ\,1720, and $\sim$3 times higher than RXJ~2129. This greatly affects the quantity of radio detections particularly the low-luminosity radio sources (L$_{\rm 1.4\,GHz}{<}10^{23}$\,W\,Hz$^{-1}$) which likely belong to the 'S' source category and whose radio emission is probably linked to star formation.

\subsubsection{Optical counterparts of the radio sources}
While our clusters do contain diffuse radio sources such as radio haloes and mini-haloes, we are primarily interested in sources associated with individual non-central galaxies, with radio emission due to star formation or AGN. We crossmatch our detected radio sources with the detailed redshift catalogues for each cluster, to unambiguously find instances of radio emission associated with member galaxies.

\subsection{Computing the radio power from non-central sources}

We compute the radio power of each source based on its single-band 610~MHz flux. We assume that all sources have a power law spectrum described by S$_\nu \propto \nu^\alpha$, where S$_\nu$ is the flux density at frequency $\nu$ and $\alpha$ is the spectral index. We adopt a value of $\alpha$=-0.8, typical for synchrotron emission from radio galaxies at frequencies $\sim$1~GHz \citep[e.g.,][]{Condonetal92,KellermannOwen98}. For convenience of comparison with previous works, we calculate the radio power at 1.4~GHz using the following equation:

\begin{equation}
\label{eq:power}
{\rm L}_{\nu} = 4\pi {\rm D_L}^{2} (1 + z)^{-(\alpha + 1)} {\rm S}_{\nu},
\end{equation}

\noindent where L$_{\nu}$ is the radio power in W Hz$^{-1}$ at the frequency $\nu$, $D_L$ is the luminosity distance and $z$ is the redshift. We use the terms radio power and radio luminosity synonymously in the text.

We note that while our adopted spectral index should be representative of most sources in clusters, the actual spectral index will vary from source to source. Differences in spectral index of $\pm$0.2 would produce differences in L$_{1.4~GHz}$ of $\sim$5-10 per cent, comparable to the uncertainties on the 610~MHz flux densities.

\subsection{Volume estimation for individual non-central galaxies}
\label{subsection:volume estimation}
As mentioned previously in \S~\ref{subsection:pybdsf}, detected radio sources, based on their sizes can be broadly classified in two categories, single Gaussian (S) sources which are unresolved and Multi-Gaussian (C or M) sources which are the extended sources. However it is possible that some marginally extended sources, identified as extended by {\sc PyBDSF}, may in fact be unresolved. Such sources could be small radio galaxies with angular sizes comparable to the restoring beam that are not properly resolved, but it is also possible that small errors in the estimation of the beam shape or contributions from structured noise features could make a truly compact point source appear larger than the beam, leading to their false identification as extended. The status of these ambiguous marginally extended (C or M) sources is not entirely clear and hence we consider them as a separate class for the rest of our analysis. In order to distinguish these sources from the point and extended sources, we term these as "point-like" sources as they are classified by {\sc PyBDSF} as extended, but on examination appear almost consistent with point sources. Thus the non-central galaxies based on their linear sizes can be "point", "point-like" or "extended".

We detect a total of thirteen `C' or `M' type radio sources. Of these detected sources, two are central radio sources. The remaining non-central sources comprise seven extended and four point-like radio sources. Fig.~\ref{fig:v_sigma_r_r200} shows the phase-space distribution of all cluster members along with the detected radio sources. Point sources ('S' type) are marked by green diamonds, point-like sources are red triangles, and extended sources by orange squares. While virialized populations are strongly concentrated in the cluster core (r<0.5$\times$r$_{200}$) and at low velocity offsets (|v-cz|/$\sigma$<1), infalling galaxies are more uniformly distributed in phase-space, over -2<(v-cz)/$\sigma$<2 and out well beyond the virial radius. In general, it appears that the 'S' type radio sources are mostly present in the infalling population rather than forming a part of the virialised population since point radio sources do not congregate in cluster cores or at low velocity offsets. 86 per cent of our detected radio sources are 'S' type. The extended radio sources instead form a part of the virialized population, having low velocity offsets and lying mostly within r$_{200}$ (with the exception of one source in A1914).

The jet power of central radio sources is usually estimated based on the enthalpy of the radio lobes (itself estimated from the mechanical work required to inflate the lobes against the pressure of the surrounding ICM) divided by the timescale of their formation. This is often referred to as the cavity power, since the volume of the lobes is usually estimated based on the size of decrements in X-ray emission, assuming these to be 3D cavities, usually with an ellipsoidal shape. Since for non-central galaxies cavities may be smaller and further from the cluster centre, and therefore difficult to detect in the X-ray, we make our volume estimates from the radio images. However, we note that in some cases the volume is likely to be an overestimate. For simple, well-resolved lobe sources, the radio maps are likely to give a very accurate estimate of lobe shape. However, for sources on scales similar to the restoring beam, or for tailed sources where the radio jets may not inflate simple, single cavities, the extent of the radio emission may give a biased impression of cavity size.

We obtain the extent of lobed and tailed radio sources by decomposing them into simple geometrical shapes. A head-tail galaxy for example can be approximated by a circle  (sphere) covering the head and narrow rectangle (cylinder) covering the tail. Each lobe of an FR-I or FR-II is approximated by an ellipse which we assume to be a projected ellipsoid with rotational symmetry around the jet axis. We compute the volume and add the individual components to obtain a total volume. In general we conservatively treat the line of sight depth of most components as being equal to the shorter of the two projected dimensions. This means that, e.g., an elliptical region is assumed to be a prolate  ellipsoid in projection, rather than an oblate ellipsoid viewed from the side.

For point or point-like sources, we are limited by the resolution of the radio data, and while most are likely truly compact, some could in fact host small-scale radio jets and lobes. Such sources could be inflating cavities in the ICM, and causing heating. To take an extreme example, we could imagine young, radio luminous, over-pressured jet sources inflating a radio cocoon and expelling much of the ICM gas in their vicinity. Such sources would inject energy into the ICM through cavity inflation but also through shocks, and so even assuming a cavity the size of the radio source would underestimate energy input. Indeed it must be borne in mind that almost all jet sources, central or non-central, will at some stage expand supersonically and drive ICM shocks. It is rare that the available X-ray data allows shock strengths to be measured, and certainly for our sources it is not possible. We must therefore follow the approach of most previous studies and set aside the possible contribution from shocks. For the point-like sources, we therefore estimate an upper limit on the volume and jet power, on the basis that an ellipsoidal cavity with the projected size equal to that of the radio emission (i.e., for point sources, the beam size) could be present.

\section{X-ray Analysis}
\label{x-ray data analysis}
\subsection{X-ray data reduction}
\chandra\ data reduction was carried out using \textsc{ciao} 4.11 and CALDB 4.8.4 following standard procedure as prescribed in the \chandra{} analysis threads \footnote{\url{http://asc.harvard.edu/ciao/threads/index.html}}. Observations of the target clusters using the Adavanced CCD Imaging Spectrometer (ACIS) were identified in the \chandra\ archive. The data for each observation were first re-processed using the task {\sc chandra\_repro}, and new bad pixel files were created. Cosmic-ray after glow correction was also applied. After removal of bright point sources, periods associated with background flaring events were identified using the task {\sc deflare} with the {\sc clean} method, and removed from the event lists. We combined multiple reprocessed event files to obtain a re-projected and merged event file using the task {\sc merge\_obs}. The merged event file provided the greatest sensitivity for detecting faint point sources. The PSF maps were combined from the individual observations. Point sources were identified by the task {\sc wavdetect} and removed from the diffuse emission originating from the cluster's ICM. Sources associated with the diffuse emission or potential AGN of the cluster members were retained.
The matching ACIS background files ("blank sky") were identified using the task {\sc acis\_bkgrnd\_lookup}. In the case of multi-chip observations, the {\sc acis\_bkgrnd\_lookup} task returns a list of datasets. We merged all the blank sky files mentioned in the list using the task {\sc dmmerge}. This gave the reprojected blank-sky background files for diffuse emission analysis and spectral extraction. The background data were renormalised to match the 9.0-12.0~keV count rate of the corresponding source observations. We summarise the X-ray observations used in our analysis in Table.~\ref{table:X-ray_obs}.

\subsection{X-ray spectral analysis}
\label{subsection:spectral analysis}
Spectra were extracted using the {\sc specextract} task from both the event and background files in circular annuli. The annuli were chosen to achieve a minimum signal-to-noise ratio (SNR) of 60. For each cluster, a deprojection was carried out in \texttt{Xspec} 12.9.1p using the {\sc projct*phabs*apec} model, to produce 3D deprojected profiles of gas parameters including temperature, metal abundance and electron number density. From these parameters we calculated the entropy, pressure, luminosity, and cooling time in the 3D radial bins. We defined the cooling radius as the radius within which the cooling time was less than 7.7 Gyr, the approximate look-back time since $z=1$. This is the typical age of the clusters since their formation. Thus the cooling radius is the boundary within which the gas could have reasonably cooled within the lifetime of the clusters.

The analysis carried out for Abell\,1758 was different from the rest as this is the only cluster undergoing a major merger at the present epoch with clearly separated cores. Abell~1758 actually consists of four sub-clusters, with two sets of merging pairs. We used the northern pair, analysing the NW and NE sub-clusters separately. In each analysis, emission from the other sub-cluster (and the southern pair) was excluded, to avoid any contamination. Thus we obtained a set of parameters for each sub-cluster.

\subsection{Pressure estimation for individual non-central galaxies}
\label{subsection:pressure estimation}

We determined the ICM pressure at the location of each radio source. For all sources pressures were estimated using an extrapolated pressure profile. We fitted the universal pressure profile of \citet{Arnaudetal10} to our deprojected pressure profiles, adopting the optical redshift of the cluster. The only fitting parameter was the cluster mass, effectively the normalisation of the model. Pressures were then calculated based on the fitted profile, at the position of each radio source.

In estimating the ICM pressures, we assume that the projected distance of each source from the cluster centre is a good estimate of its true radius within the cluster. We will discuss this, and other potential biasing factors associated with this approach in more detail in \S\ref{subsection:biases}.

\input{tables/X-ray_observations.tex}
\input{tables/xray_properties_galaxies.tex}

\section{Results and Discussions}
\label{results}

\subsection{Radio populations of the clusters}
While all our non-central radio galaxies are selected to have redshifts consistent with cluster membership, many of the extended sources also have radio morphologies characteristic of interaction with an ICM, e.g., head-tail sources, narrow angle tail (NAT) and wide angle tail sources (WATs). For marginally resolved point-like sources, there is always the possibility that some small number may be interlopers, particularly those with high velocities relative to the cluster mean. All the extended radio sources in our sample are hosted by galaxies with an early-type morphology. Relevant details of the clusters are described in Appendix~\ref{appendix:cluster notes}, images and properties of the non-central radio sources can be found in Appendices~\ref{appendix:images} and \ref{appendix:table} respectively. We briefly describe here the non-central radio sources in each cluster, as identified by \textsc{PyBDSF}.

Abell\,383 has a FR-I galaxy located to the east of the cluster centre (ID: 15; see Table~\ref{tab:x-raytable}). In Abell\,1758, we find a head-tail galaxy (ID: 23) in the far west corner of the NW sub-cluster and a blob-like marginally extended source (ID: 26) located north of the BCG of the NW sub-cluster. After examination, we categorize the latter as a point-like source.

In RXJ~1720, there are three extended galaxies, one head-tail (ID: 63) which is close to the cluster core, a WAT radio galaxy (ID: 64) located south-east of the cluster centre, and another blob-like source (ID: 76) located towards the north of the cluster. Again, after examination this third source is recategorized as a point-like source.
In RXJ\,2129, we find a prominent WAT (ID: 84) located in the north-east part of the cluster. ZwCl\,3146 has one extended FR-I source (ID: 91) south of the cluster centre and another (ID: 92) point-like source located to the east. Abell\,1914 has two sources, one (ID: 33) with a head-tail morphology, the other point-like (ID: 38). Both sources are located far from the centre of the cluster and thus do not appear in the limited field of view of the Appendix figure.
Abell\,1914 \citep[e.g.][]{Mandaletal19} and Abell\,1758 \citep[e.g.][]{Schellenbergeretal19} both host radio halos associated with their ongoing mergers. Abell\,1914 also hosts a radio phoenix. The other four clusters have a relatively relaxed X-ray morphology \citep{Owersetal11,Mantzetal14,Kaleetal15b,Savinietal19,ZuHoneSims19} and at least three of them host mini-haloes \citep[RXJ~1720, RXJ~2129, ZwCl~3146,][]{Giacintuccietal14a,Kaleetal15b,Giacintuccietal19}, which are usually found in dynamically relaxed cool-core clusters.

\subsection{Heating vs Cooling}
\label{subsection:AGN feedback}

\begin{figure*}
    \centering
    \hspace*{-0.5cm}
    \includegraphics[width=\textwidth]{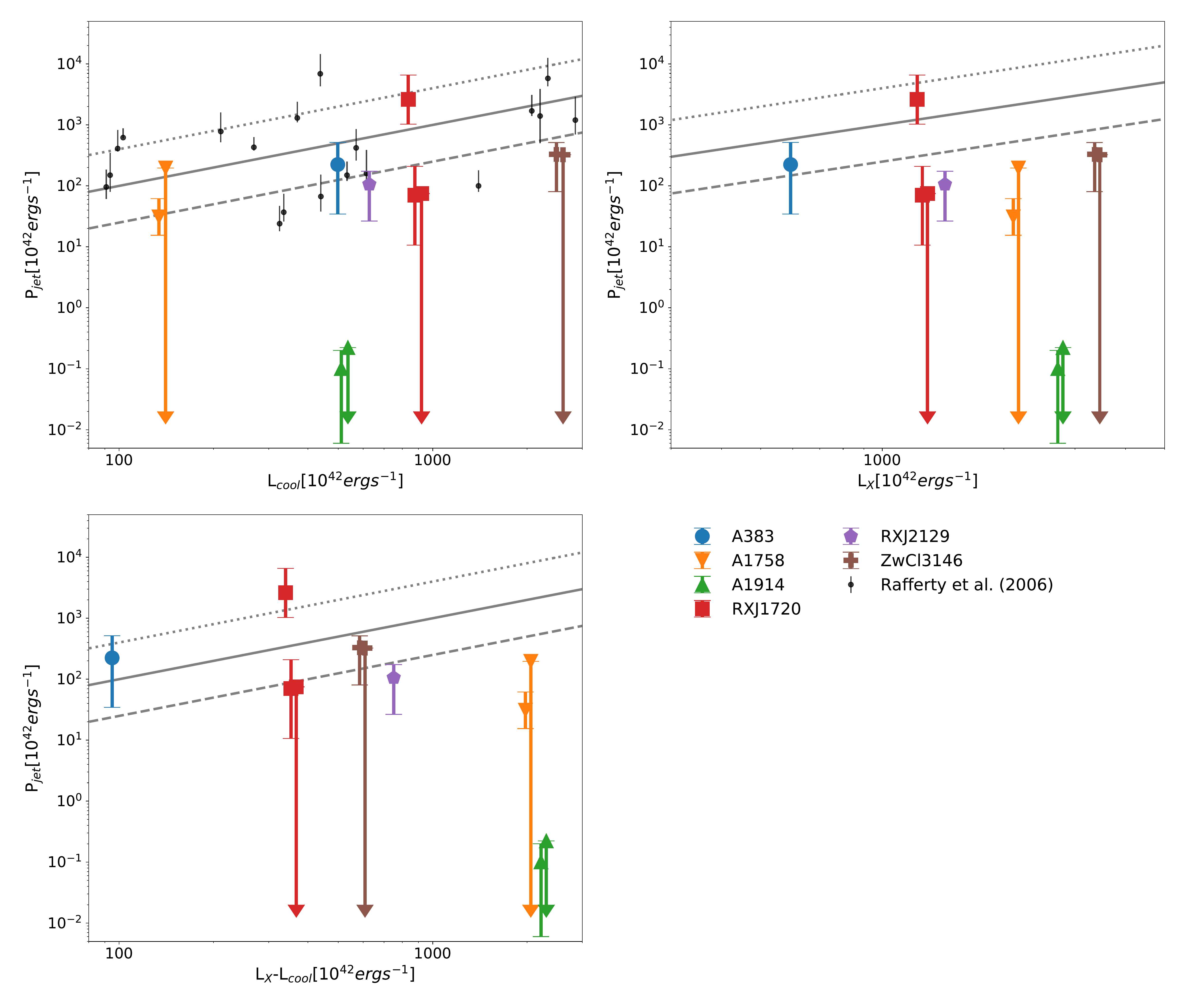}
    \caption{These plots show the jet power P$_{\rm jet}$ associated with the individual non-central galaxies vs the values of  L$_{\rm{cool}}$, L$_{x}$, and L$_{x}$ - L$_{\rm{cool}}$ of their host clusters. Points with error-bars represent the extended sources and upper-limits denote the marginally extended point-like sources (see \S{\ref{subsection:pybdsf}} for details). X-axis value is slightly shifted for galaxies in the same cluster for clarity. In each panel, The solid line indicates equality between P$_{\rm jet}$ and the luminosity, while the dashed and dotted lines are a factor of 4 lower and higher respectively.}
    \label{fig:pcav_lx_lcool}
\end{figure*}

The enthalpy of a radio lobe inflated ICM cavity is defined as 4$p_{\rm th}V$ where $p_{\rm th}$ is the ICM thermal pressure and $V$ is the lobe (or cavity) volume. We define the jet power (P$_{\rm jet}$, i.e., the power required to inflate the lobe) as:

\begin{equation}
\label{equation:PcavX}
{\mathrm{P}}_{\mathrm{jet}}=\frac{4p_{\rm th}V}{t_{\rm sonic}}
\end{equation}

\noindent where $t_{\rm sonic}$ is the sonic expansion time, i.e., the time taken for the tip of the radio jet to reach its current distance from the core. This can be calculated from:

\begin{equation}
\label{eq:tsonic}
{\mathrm{t}}_{\mathrm{sonic}}=\frac{r}{\sqrt{\frac{\gamma kT}{m}}},
\end{equation}

where $r$ is the mean distance between the central nuclei and the edge of the source, $\gamma$ is the adiabatic index of the ICM (taken to be 5/3), $k$ is Boltzmann's constant, $T$ is the ICM temperature, and $m$ is the mean particle mass in the ICM (0.6$\times$ the proton mass). Where galaxies fall within our radial spectral profiles, we use the deprojected temperature of the bin which encompasses their projected radius. For galaxies at larger radii, we use the deprojected temperature of the outermost bin of the radial profile, noting that the outer gradient of cluster temperature profiles tend to vary only slowly with radius \citep[e.g.,][]{LeccardiMolendi08}. Various derived parameters for each extended or point-like radio source including the global cluster parameters such as the cooling radius (r$_{\rm{cool}}$) and luminosity (L$_{\rm{cool}}$) are listed in Table.~\ref{tab:x-raytable}.

The three plots in Fig.~\ref{fig:pcav_lx_lcool} show the relation between the jet power and the cooling luminosity, L$_{\rm{cool}}$, the total bolometric luminosity of the cluster, L$_{x}$ \citep[taken from][]{Cavagnoloetal09}, and the luminosity of the ICM outside the cooling region, L$_{x}$-L$_{\rm{cool}}$. Sources identified by \textsc{PyBDSF} as extended are marked by coloured  points. For multiple radio galaxies in the same cluster, the three ordinates (L$_{\rm{cool}}$, L$_{x}$, L$_{x}$-L$_{\rm{cool}}$) have been shifted apart slightly on the x-axis for visibility. For comparison, we have overplotted cluster-central  radio galaxies from \citet{Raffertyetal06}, with each point representing the total jet power (which may include contributions from multiple cavities) calculated using the buoyancy timescale. Buoyancy timescales are generally longer than the sonic timescale we have adopted for our non-central radio galaxies, and are likely most accurate for older, passively evolving cavities, whereas sonic timescales will be more accurate for younger cavities still being inflated by radio jets. The Rafferty data points are included to show the degree of scatter in the correlation between central galaxy jet power and cooling luminosity, which is quite large, perhaps two orders of magnitude.

Each cluster is represented by a different colour and marker. Uncertainties are indicated by error bars for the jet sources (lobed sources, head-tail, NATs and WATs). Point-like sources are considered as upper limits, since their apparent size (driven by the restoring beam of the radio data) is likely an overestimate of their true volume. However, these upper limits can be quite large, if the sources are near the core of their clusters, where pressures are high.

As shown by the Rafferty et al. points, central radio galaxies are generally capable of producing cavities with enough power to balance radiative losses from the cooling region, though there is considerable scatter in the relation. Perhaps surprisingly, our non-central galaxies appear to be capable of producing comparable levels of energy injection, even though some of our sources (e.g., those in Abell 1914) fall well below the relation. This suggests that, at least to a first approximation, non-central galaxies can potentially play a significant role in heating the ICM.

Comparison with the total luminosity and luminosity outside the cooling region is relevant for our sources, since they often lie outside the cooling radius of their clusters. Some of our sources are relatively powerful, but in the outer part of the cluster, where they are likely to affect parts of the ICM which already have long cooling times.

While the jet powers we estimate for the non-central sources are comparable to the luminosities of their clusters, our data do not show a clear correlation between P$_{\rm jet}$ and L$_{\rm x}$ or L$_{\rm cool}$. If non-central sources were playing a critical part in the regulation of cooling in the central regions of their clusters, such a correlation would be required. Our limited sample size makes identifying such a correlation difficult; a larger sample might still show a correlation between P$_{\rm jet}$ and L$_{\rm cool}$. But from the evidence we have, it seems that non-central sources can have a significant heating impact on their surroundings, but that cluster-central sources are the dominant factor in balancing ICM cooling over the long term.

Statistically we find 1.2 non-central extended radio sources per cluster, suggesting that at any given time at least one source is active. It therefore seems likely that many clusters will have energetically important non-central radio galaxies, and that at any given time those clusters are likely to have at least one such galaxy making contributions to heating the ICM.

\subsubsection{Contribution from radio point sources}

As noted in \S\ref{subsection:volume estimation}, given the resolution of our radio data, the true source of emission in radio point sources is uncertain. Many or all may be dominated by emission from star formation or AGN accretion disks, but it is possible that some host small-scale jets and lobes which we are unable to resolve. We therefore take a conservative approach, calculating upper limits on the enthalpy of individual sources and summing them to determine whether the point source population as a whole could have a significant heating effect on the ICM.

We can estimate upper limits on the jet power of such sources by assuming that they have inflated a cavity of size equal to the GMRT 610~MHz beam (6$^{\prime\prime}$) which at the mean redshift of the sample corresponds to a linear size of 21.24~kpc. This is similar to the approach we use for the marginally extended point-like sources, for which we use a single ellipsoidal cavity with size determined by the radio emission, but consider the resulting volume and upper limit. We can also estimate jet power indirectly, using the empirical linear relation that connects radio luminosity and jet power \citep[e.g.][]{OSullivanetal11b}, though the large scatter in the relation makes this method more uncertain. Both methods suggest low jet powers for the point sources, with minimal impact to the ICM even summing the power output of the population. Taking into consideration the likelihood that most of the point source population probably consists of SF dominated sources, and will include some AGN that are not currently launching jets, we do not consider the energetic contribution of this population further.

\subsection{Radio power output for the non-central galaxies}

\begin{figure*}
\centering{
\includegraphics[width=\textwidth]{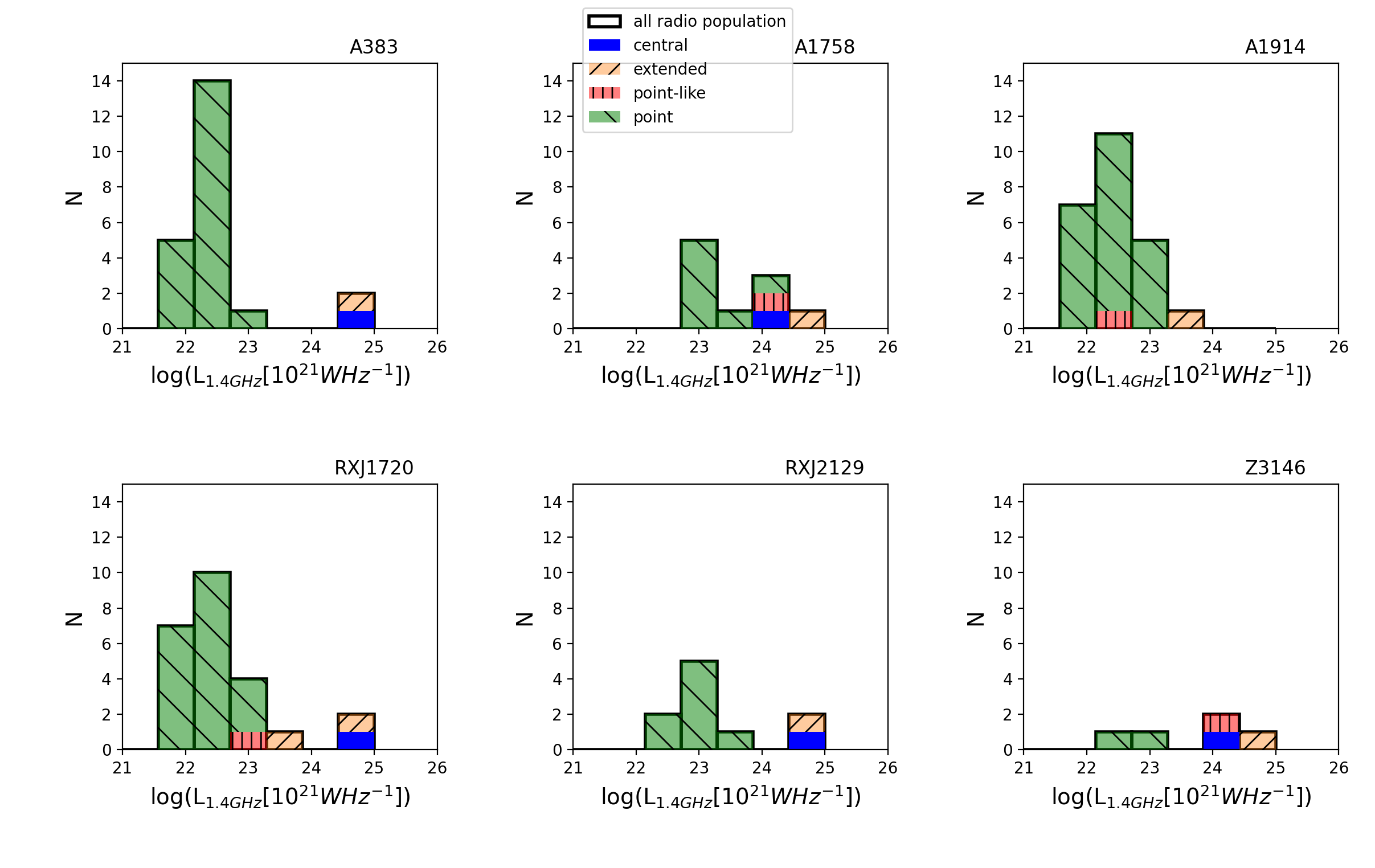}
}
\vspace{-1cm}
\caption{Distribution of the radio power at $1.4$ GHz for the clusters in our sample, Top panel: clusters Abell\,383, Abell\,1758 and Abell\,1914. Bottom panel: RXJ\,1720, RXJ\,2129 and ZwCl\,3146. The radio luminosity function corresponding to the radio detected galaxies for point (green hatched), point-like (red vertical), extended (orange hatched) and central (blue plain) sources are shown. Both the extended and central sources are powerful as compared to their point source counterparts. The point-like sources show a large scatter in their radio power. Radio powers for the central sources can be contaminated by emission from the radio halos and mini-halos (see text for details).}
\label{fig:Radio_Power}
\end{figure*}

In Fig.~\ref{fig:Radio_Power} we show the distribution of radio luminosity at 1.4 GHz for all non-central sources (computed using Equation.~\ref{eq:power}) and compare it to their respective central sources.

The mean and deviation of log(L$_{\rm 1.4\,GHz}$) for the four types of radio sources, namely the non-central point, point-like, extended and the central sources are 22.5$\pm$0.5, 23.5$\pm$0.9, 24.4$\pm$0.5 and 24.6$\pm$0.2 respectively.

In general for the non-central sources, extended sources are always the brightest, exceeding the radio luminosity of all the point sources by at least a couple of orders of magnitude. Point-like sources fall between the two categories and can be quite bright. Cluster central sources are generally very bright, but the non-central extended sources or point-like sources can be as bright or brighter. It is notable that the mean brightness of cluster central sources and non-central extended sources are identical within the uncertainties.

The central sources, in some cases, are very luminous but not extended in our data. The extra flux in central sources appears due to the presence of a non-AGN component, usually a mini halo, which we resolve out. In some cases, however, the host galaxy could not be separated from the additional component and hence remains unresolved (see Appendix \ref{appendix:cluster notes} for more details). Thus high radio brightness in central sources can be attributed in part to the shocks and turbulence in the ICM which ignites the relativistic particles present in the cluster core.

\begin{figure*}
\centering{
\includegraphics[scale=0.85]{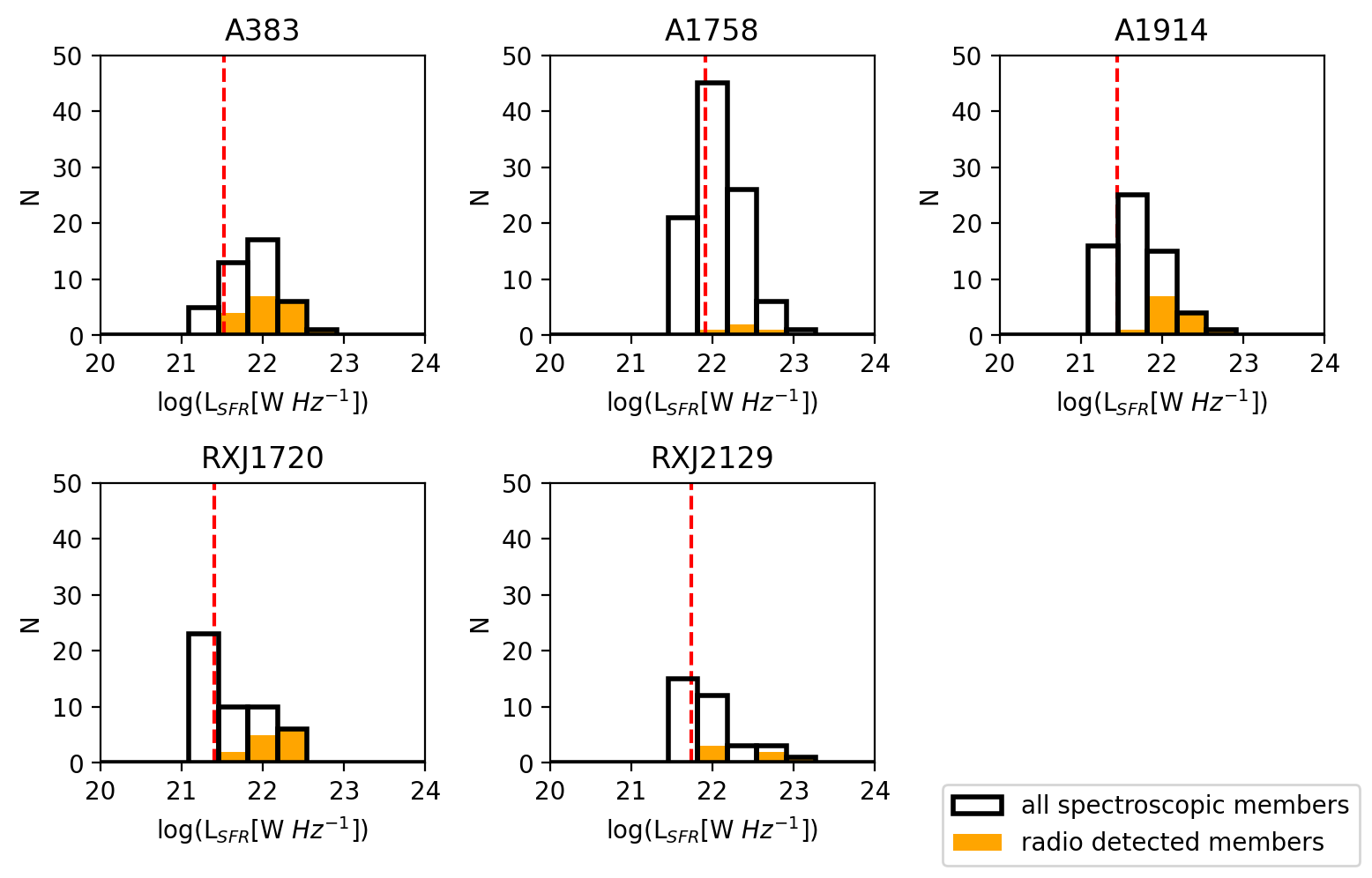}
}
\vspace{-2mm}

\caption{Distribution of the expected $1.4$ GHz radio luminosity arising from star-formation for five clusters (black unfilled). The order of the plots is as in Fig.~\ref{fig:Radio_Power}, with ZwCl~3146 excluded. The star-formation rates were computed from the 24 $\mu$m band of Spitzer/ MIPS converted to their radio equivalent luminosity using the relation of \citet{Hainesetal11}. The red dotted lines indicate the completeness limit corresponding to the flux limit of 0.4 mJy in the 24 $\mu$m band. The distribution of star-formation luminosity (\textsc{L$_{SFR}$}) corresponding to the radio detected galaxies (orange filled) is overlayed.}
\label{fig:SF_luminosity}
\end{figure*}

\subsection{Star formation in the radio-detected population}

Given the sensitivity of our GMRT observations, we can expect to detect radio continuum emission from star formation in some of our galaxies as well as from AGN. While the morphology of well-resolved extended sources usually confirms AGN as the dominant emission source, the relatively weak point sources in our sample are more likely to have contamination from star formation. Hence we compare galaxies detected in the radio in our sample with available star-formation rates (SFRs) estimated from other bands to understand the dominant source of radio emission. \S\ref{sec:SFR} describes the SFRs used for our cluster galaxies. No SFRs were available for ZwCl~3146, which we therefore exclude from all analyses where SFR is required.

Fig.~\ref{fig:SF_luminosity} shows the distribution of the expected 1.4~GHz radio luminosity from star-formation (L$_{SFR}$) in the five clusters, using the conversion of \citet{Hainesetal11}. The black unfilled histogram shows the distribution of all the galaxies whose SFR is estimated using the available 24$\mu$m MIPS data. The orange filled histogram shows the radio detected galaxies. The radio completeness limit corresponding to the limit of the SFR sensitivity in the 24$\mu$m band is shown by the red dotted line on the left for each cluster. We observe that the galaxies detected in radio preferentially have a higher SFR than the ones not detected in radio. From the plot, it seems that the MIPS data is more sensitive to star formation than the GMRT data, however the low luminosity radio detections depends strongly on the sensitivity of the GMRT data. The difference in sensitivity of the radio maps for the different clusters leads to A1758 having far fewer radio detections at the same SFRs than the other clusters. We see that the clusters A383, A1914 and RXJ1720 appear complete at high SFRs: SFR>4$\M{}$yr$^{-1}$, or log(L$_{1.4GHz}$)>\textasciitilde{}22.2 whereas in case of A1758 we detect only the most actively star-forming galaxy (SFR=20$\M{}$yr$^{-1}$) in radio. We note that a few galaxies are also missing from the sample as the they fall outside the MIPS coverage.

\begin{figure*}
    \centering
    \includegraphics[width=\textwidth]{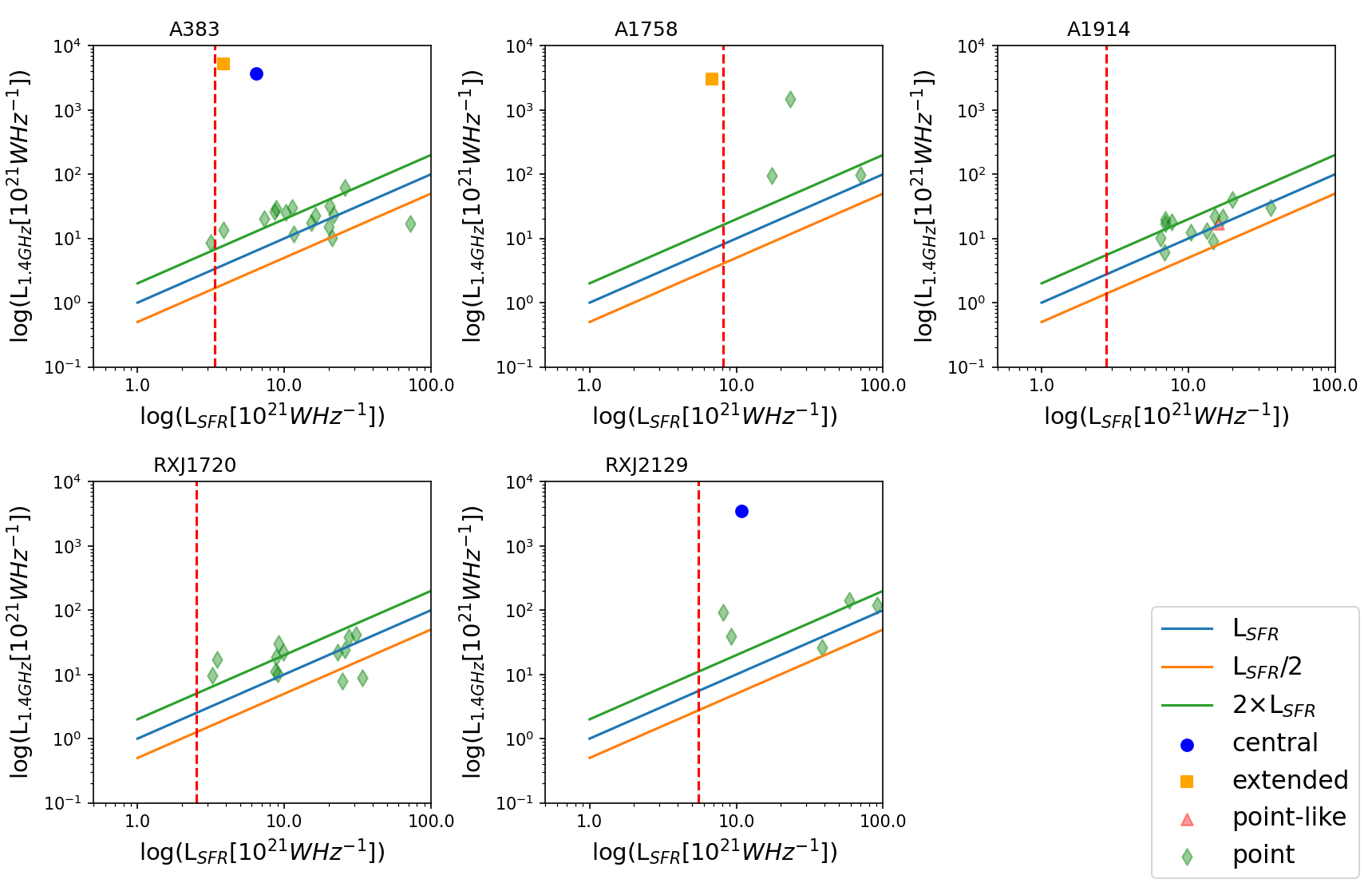}
    \caption{Estimated 1.4~GHz radio luminosity (L$_{1.4 GHz}$, based on our GMRT 610~MHz measurements) vs predicted 1.4~GHz luminosity from star formation \citep[L$_{SFR}$, using the relation of][]{Hainesetal11} for the non-central galaxies of our cluster sample, except ZwCl\,3146. Panels are in the same sequence as in Fig.~\ref{fig:Radio_Power}. In each panel the three lines represent L$_{1.4 GHz}$=L$_{SFR}$, L$_{1.4 GHz}$=2$\times$L$_{SFR}$, and L$_{1.4 GHz}$=0.5$\times$L$_{SFR}$.}
    \label{fig:l1.4_lsfr}
\end{figure*}

We further determine which of our radio detected galaxies may have significant radio flux arising from star formation, from Fig.~\ref{fig:l1.4_lsfr} which represents the radio luminosity at 1.4~GHz as a function of the equivalent radio luminosity expected from star formation.

We find for a few extended or central sources, whose star-formation rate could be estimated, the measured radio luminosity exceeds the contribution expected from star formation by a large factor, generally two orders of magnitude, but most point-sources lie close to the equivalence relation, suggesting most of their radio emission originates from star-formation. More than half of the point radio source population whose star-formation rates could be estimated have more than 50\% of their radio flux contributed by star formation, and almost one-fourth of the radio detected population are likely star-forming dominated, with IR-predicted SFRs suggesting radio luminosities that would exceed their measured 1.4 GHz luminosity (see Table.~\ref{tab:D1_SFR}).

\begin{figure*}
\centering{
\includegraphics[width=\textwidth]{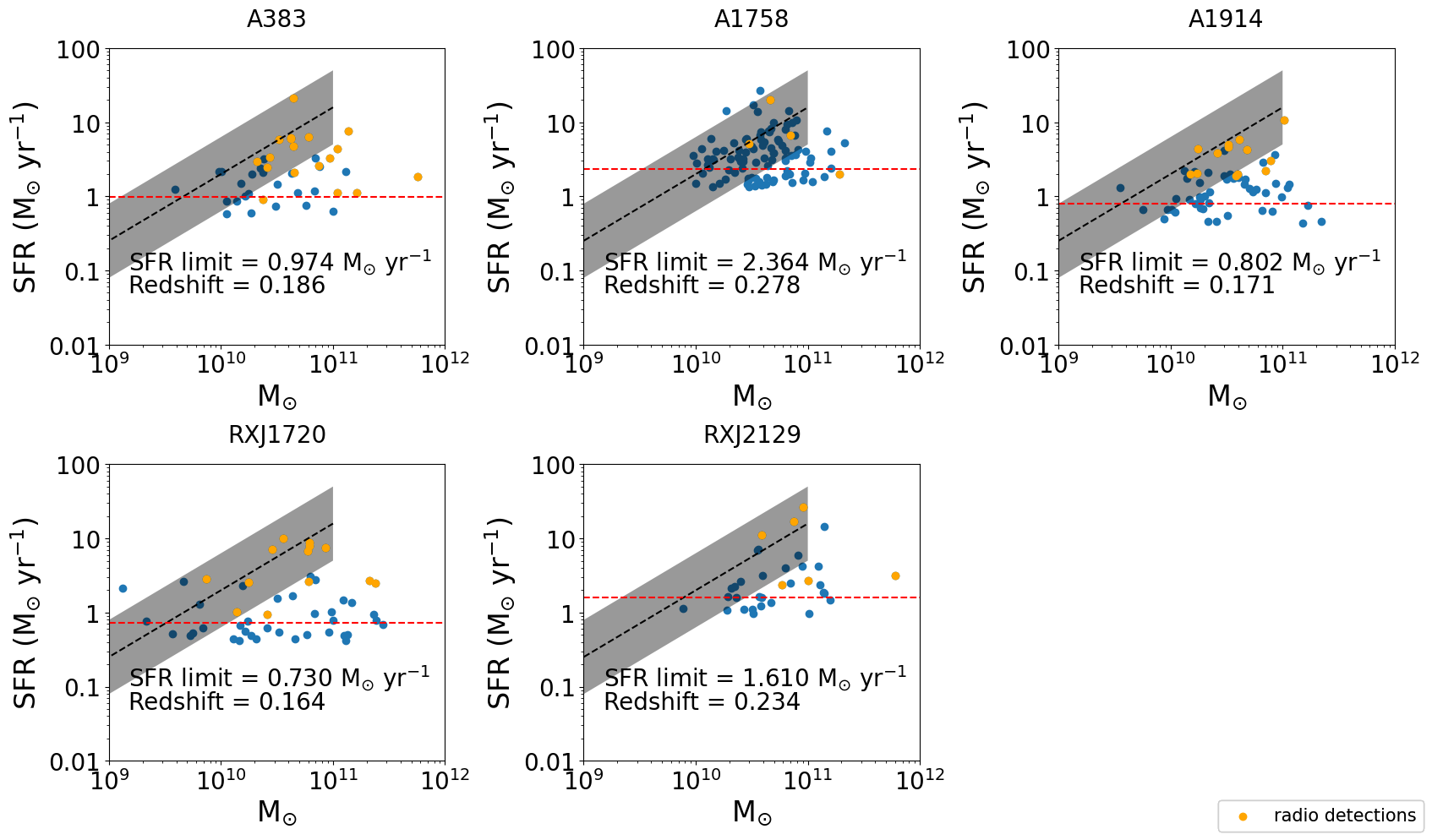}
}
\caption{The SFR- M$_{*}$ relation for all the galaxies having an SFR measured from the 24 $\mu$m flux for our cluster sample except ZwCl\,3146, in the same sequence as Fig.~\ref{fig:Radio_Power}. The red dotted line represents the completeness limit for the star-formation for each of the cluster which corresponds to a threshold flux density of 0.4 mJy. The black dotted line represents the main star-forming sequence from \citet{Wuytsetal11}. The orange points represent galaxies with detected radio point sources.}
\label{sfr-m}
\end{figure*}

In Fig.~\ref{sfr-m}, we plot the estimated SFR and stellar mass for all our galaxies. The star-forming main sequence is represented by an empirical relation taken from \citet{Wuytsetal11}. The points represent cluster members with measured SFR, with orange points representing radio-detected systems. We observe that a significant fraction (about 30-60\%) of galaxies detected in the radio lie on the star-forming main sequence, indicating a composite population with both star-formation and nuclear activity.

In summary, we find that a significant fraction of the radio-detected galaxies in our clusters, primarily those identified as radio point sources, have fluxes consistent with arising at least partly from star-formation emission. Most of these systems are relatively massive galaxies, and many fall on or close to the star-forming main sequence.

\subsection{Optical-NIR properties of the host galaxies}
\label{subsection:kband host magnitude}
\begin{figure*}
    \centering
    \includegraphics[width=\textwidth]{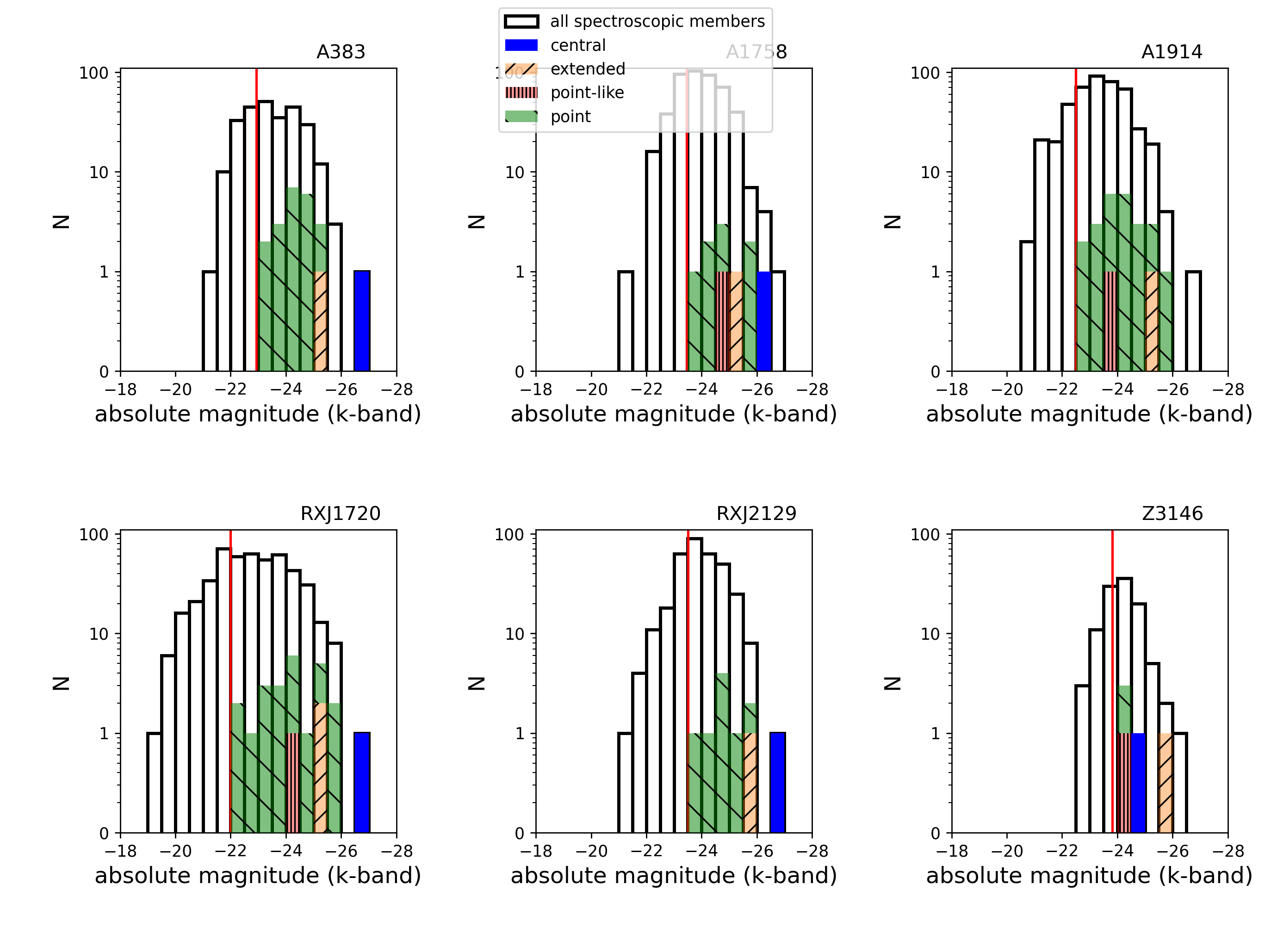}
    \caption{Distribution of k-band magnitudes for all available spectroscopic members for our cluster sample in the same sequence as Fig.~\ref{fig:Radio_Power}. The ones detected in radio have been colour coded according to our classification scheme. The vertical red line in the plot demarcates the non-detections in radio consisting of the bottom 20 percent of galaxy population in case of Abell\,1914 and bottom 30 percent for the rest of the sample.}
    \label{fig:kmag}
\end{figure*}

Galaxies hosting point radio sources can also be distinguished from galaxies hosting extended radio counterparts in their optical-NIR properties. Fig.~\ref{fig:kmag} shows the distribution of $K$-band galaxy absolute magnitudes for the cluster members drawn from UKIRT/ WFCAM imaging. The red line drawn in the figure depicts the threshold magnitude of the host galaxies above which the radio emission is detected.
 
The radio detected sources of all types tend to fall at the upper end of the overall distribution of galaxy magnitudes, with M$_{K}{<}$-22.0. This demonstrates that we lack the radio sensitivities to detect low-level star formation in the smaller systems. As expected, cluster-central radio sources (where present) are hosted by the brightest cluster galaxies, as defined by their $K$-band luminosity.
Galaxies hosting extended radio sources tend to fall at the upper end of the distribution, with all extended radio sources located within the most luminous $\sim$5\% of galaxies, those with M${_K}$<-25.0 (corresponding to a stellar mass limit of 10$^{11}$ \M{}).
It is well known that large-scale radio jet sources tend to be hosted by the most massive early-type galaxies \citep[e.g.][]{Hickoxetal09}, and this distribution suggests that while some of our radio point sources may be capable of growing into extended sources in future, some are likely hosted by galaxies which are too small to produce such high power sources. The correlation between the supermassive black hole mass and the mass of the host galaxy \citep[e.g.][]{Ferrarese02, Bandaraetal09} is well established. Since $K$-band magnitudes of these galaxies are a proxy of the underlying galaxy mass, only galaxies above a certain threshold in their $K$-band magnitudes are likely to be capable of hosting a supermassive black hole that can produce such powerful radio sources. For our flux limited sample, around 7\% of these most massive galaxies are detected in the radio.


\subsection{Detection statistics}
 
Our analysis finds a overall radio detection rate of $\sim$4.5\% of the galaxy population across our six clusters. However the extended sources, which appear to be the dominant source of heating, are rarer. We detect a total of ninety-five radio sources in our six clusters, which on average gives sixteen radio sources per cluster. We detect roughly ten times as many point sources as we do extended sources.

\begin{figure}
\centering{
\includegraphics[width=\columnwidth]{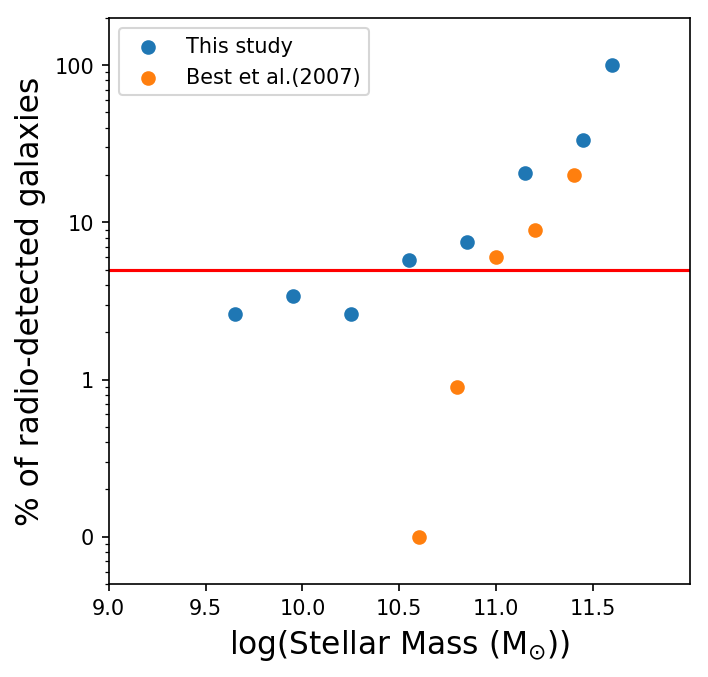}
}
\caption{Distribution of the radio-loud fraction in our sample (blue) and the sample of Best et al. (2007, orange). The sensitivity of GMRT results in a higher radio-fraction at both lower and higher stellar mass haloes. The detection threshold of the Best et al. catalogue is at 10$^{10.5}$ \M{}. The red line represents the overall mean detection rate of our sample.}
\label{fig:Best2007}
\end{figure}

In order to compare the detected radio sources in our GMRT analysis to those of NVSS and FIRST, we crossmatched our spectroscopic sample with the radio catalogue provided by \citet{BestHeckman12}, we get a total of seven radio sources, out of which six are classified as AGN, mostly from the non-central galaxies, amounting to one radio source per cluster. Only the most prominent radio sources are detected by FIRST/ NVSS.
In Fig.~\ref{fig:Best2007}, we present a comparison of our radio detected fraction to that of \citet{Bestetal07} as a function of stellar mass. As expected, the figure shows that our deeper GMRT data traces radio emission down to galaxies of lower stellar mass. \citet{Bestetal07} detects only radio sources associated with galaxies above a stellar mass of $10^{10.5}$\M{}. Nonetheless, both samples show the same increase in the fraction of radio detections with increasing stellar mass above this limit.

As has already been pointed out, these prominent radio sources are also the most capable of injecting energy into the ICM. The fainter point radio sources are in many cases likely dominated by star formation, and in any case are unlikely to make a significant contribution to ICM heating. This implies that it may be possible to estimate the heating caused by these prominent non-central radio sources across wider available group and cluster catalogues using only NVSS and FIRST; deeper radio observationa may not be required. Detection statistics for individual clusters are provided in Table.~\ref{tab:detection_statistics}.

\input{tables/detection_statistics}

\subsection{Potential sources of bias}
\label{subsection:biases}
Having described our results, it is important to consider potential biases which might affect them, in particular the uncertainties in determining jet power, P$_{\rm jet}$, for the extended radio sources. Since P$_{\rm jet}$ is dependent on the cavity volume, the surrounding pressure and the expansion timescale of the source, uncertainties arising from each parameter would contribute to the total uncertainty.

In contrast to previous studies of cluster-central sources, we are unable to confirm the presence of cavities from X-ray observations, and have to assume cavities are present and have sizes comparable to the extent of the radio emission. As described in Section~\ref{subsection:volume estimation}, we follow the usual assumption that the volumes of the radio structures can be approximated by simple goemetrical shapes, cylinders and ellipsoids. 

Since the depth of such structures along the line of sight is unknown, we include additional uncertainties in our calculation, allowing a baseline factor of 2 uncertainty in the line-of-sight dimension, and for ellipsoids, where the semi-minor axis is used to estimate the depth, allowing that it could be as large as the semi-major axis. These uncertainties are included in the estimates of jet power.

As is typical for analyses of cluster-central sources, we assume that radio sources are aligned close to the plane of the sky. If this is inaccurate, the volumes of radio structures may be underestimated, since their longest axes may appear shorter due to projection, and timescales may also be underestimated. However, since these biases would also affect the cluster-central sources we compare with, and orientation axes should be essentially random, we do not consider this an important source of bias.

The ICM pressure used in estimating the jet power for each source is estimated by extrapolating the deprojected pressure profile for each cluster to the source position. There are uncertainties associated with the deprojection analysis as well as extrapolation of the pressure profiles. The uncertainty on the normalization of the fitted profile produces an uncertainty on pressure at the projected radius of each galaxy; this is included in the final values. The pressure of the ICM surrounding the source also depends on the location along the line of sight with the assumed value to some extent representing an upper limit. We cannot know the true position of the galaxy in the cluster, but we can include at least a crude estimate of the uncertainty by assuming that the galaxy may be as far from the cluster center along the line of sight as it is in projection, i.e., its true radius could be as much as $\sqrt{2}$ times its projected radius. We calculate the pressure at this larger radius, and include the difference between this and the pressure at the projected radius in the uncertainty on the jet power.

Since the universal pressure profile is derived from a set of pressure profiles of relaxed clusters \citep[see][]{Arnaudetal10}, it may be less suitable for disturbed, merging systems. Comparing the measured pressure profile to the fitted model, we find that the ongoing merger in Abell\,1758 could cause an underestimate of the pressure at large radius beyond $500$ kpc which would affect the sole extended source present in the cluster. The other merger system, Abell\,1914, shows a good agreement between data and model, and is thus unlikely to be affected.

The third major assumption goes into the estimation of the timescale. As with many studies of cluster-central sources \citep[e.g.][]{Birzanetal04} we have used the sonic timescale as an estimate of the expansion time of the radio sources(see Eq.~\ref{eq:tsonic}).

One source of error in this estimation of timescale arises from the temperature value assumed for the surrounding ICM. This is derived from the deprojected temperature profile, but a number of assumptions go into these calculations. For galaxies with projected radii within the range of our measured temperature profiles, we use the temperature of the bin which includes the galaxy (or the mean temperature if it overlaps two annuli). This again assumes that the projected cluster-centric radius gives an accurate measure of the true position of the galaxy in the cluster. Galaxies whose true radius is significantly larger could fall in a different temperature bin. For galaxies whose projected position falls within the cool core of a cluster, this implies that their true surrounding temperature could be higher, implying an over-estimate of their expansion timescale, and thus an under-estimate of their power. Secondly, for sources lying outside the range of the measured temperature profile, we assume the temperature of the outermost bin is reasonably representative of the temperature at larger radii. This may overestimate the true temperature since cluster profiles generally decline at large radii. However, in both cases we are unlikely to be inaccurate by more than a factor of $\sim$2, given the range of temperatures involved and the typically mild gradients of temperature profiles at large radii \citep[e.g.,][]{LeccardiMolendi08}. This is comparable to the uncertainties involved in adopting the sonic timescale as an estimate of expansion time.

Most of the sources were estimated to have timescales within an order of magnitude of 10$^{7}$ years, with timescales for the extended sources usually in the range of a few 10$^{8}$ years. However, the sonic timescale can only be an approxmation, with the true timescales of individual sources being shorter or longer. Among well studied cluster-central AGN, examples of trans- or super-sonic lobe expansion have been observed in young or powerful sources, but most central FR-Is appear to be older sources expanding sub-sonically \citep{Crostonetal04}. However, when comparing non-central FR-Is and WATs to the equivalent cluster-central sources, any bias is likely to affect both populations. It should also be noted that since the sources used in the jet power estimates are generally detected at both 610~MHz and 1.4~GHz, they are more likely to be younger sources still powered by (and perhaps still over-pressured by) active radio jets. Old, inactive sources would be expected to have steep radio spectra and thus be more difficult to detect at these frequencies.

By contrast, the size of head-tail galaxies is independent of the jet speed. A galaxy falling into a cluster for the first time, and passing through the core, may be moving supersonically, with the length of detectable tail depending on the rate at which the radio emission fades once it becomes detached from the jets. As with volume, sources at large inclination angles may have underestimated timescales owing to foreshortening, and this might affect strongly bent radio sources more than those in which the lobes are rising and expanding buoyantly.
The volume and hence the enthalapy associated with the head-tail sources are likely overestimated. Cavities have not generally been detected coincident with head-tail sources, though they may be present on scales below the resolution or sensitivity of the available X-ray data. In the absence of cavities, energy may be transferred to the ICM through alternate means, e.g., particle interactions between the radio-jet plasma and the surrounding ICM \citep{Gittietal12}. Particle entrainment rates are expected to be high in case of maximum contact between jet-surface and the external medium \citep[][]{Crostonetal08}. Bending of jets due to sustained relative motion of the host galaxy with respect to the ambient ICM is another example where mechanical energy can be transferred to the medium as the work is done by the jets against the pressure exerted by the surrounding environment \citep[e.g.][]{Begelmanetal79}. However, it is unknown how much energy is transferred to the ICM by these processes.

Three out of seven extended sources studied in our limited sample are head-tails (ID: 23, 33 \& 63). However, two (23 \& 33) are located far from the cluster centre, and thus do not make large contributions to the feedback budget. If we exclude head-tail sources from our sample we would be left with with four sources which are scattered about the 4PV line, still making significant contributions to the energy balance of their host clusters. Our basic result therefore appears unchanged, with our sample suggesting that non-central radio galaxies can be significant sources of heating in galaxy clusters.

Thus considering possible uncertainties arising from estimating volume, pressure and timescales of the radio sources and adding them in quadrature gives an estimate of the total uncertainty in the jet power of the source. This may present a high bias in our sample data but is currently beyond our analysis scope. A systemic study of a larger sample of clusters may help resolve some of these biases.

\section{Conclusions}
\label{conclusions}

In this paper, we have used a small sample of six rich ($>$10$^{14}$~\M) galaxy clusters to investigate the potential role of of non-central radio galaxies in heating the ICM. We used a combination of radio, X-ray, IR, optical and UV data to determine the AGN power output and level of radio emission from star-formation in the non-central galaxies. As cavities become difficult to detect as one moves away from the centre of the cluster, we use the extent of the radio emission as an indicator of cavity volume. Our pressure values come from fitting the universal pressure profile to the deprojected pressure values obtained from the X-ray data. We also looked at the radio galaxy population in these clusters in terms of their radio morphology and luminosity, host galaxies and the origin of their radio emission and comment on how these quantities may be linked. The main conclusions from our study are as follows:
\begin{enumerate}
\item The extended non-central radio sources are on average two orders of magnitude more radio luminous than their point source counterparts. We find that their jet powers are in some cases sufficient to balance radiative losses from the cooling region of the ICM, or to cause significant heating in the outer parts of the ICM, depending on their location. The overall statistics suggest the presence of, on average, 1.2 extended non-central radio sources per cluster. This suggests that, although heating by the radio jets of the BCG are the dominant factor in balancing ICM cooling over the long term, at any given time a massive cluster is likely to have at least one non-central radio galaxy capable of injecting significant energy into the ICM.

\item  The most prominent, radio luminous sources (e.g. WATs and FR-I) in our sample are hosted in galaxies with the brightest $K$-band magnitudes and highest stellar masses. Only $\sim$7\% of the most massive galaxies host such powerful radio sources with the radio detection rate across the whole galaxy population being 0.5\%. Thus only one in fourteen galaxies above a certain $K$-band magnitude is active with prominent jets. These extended radio galaxies are generally found within the virialized zones of the clusters.

\item The majority of radio sources in our sample were unresolved, outnumbering the extended sources by an order of magnitude. These sources are characterised by low radio luminosities, relatively high star-formation rates and are mainly found in the infalling galaxy population. Radio emission in these unresolved radio galaxies primarily originates from star-formation, though AGN may contribute. These galaxies have lower stellar masses than the hosts of extended radio sources, but span a much wider mass range. A few sources have marginally extended radio emission (the point-like subset). Radio luminosity of these sources were higher than those of point sources but not as high as their extended counterparts. Although the underlying structure remains unresolved and at best slightly extended, these sources potentially have an energetically favourable output but higher resolution radio studies would be needed to determine whether they are truly extended. We thus consider the calculated energy output of these sources as upper limits. The overall radio detection fraction of point and point-like subset is 4.5\%.

\item While targeted GMRT observations are significantly more sensitive to radio emission in less massive cluster galaxies (${<}10^{10.5}$\M{}) than surveys such as NVSS \& FIRST, the additional population of galaxies detected tend to be relatively weak radio sources, whose emission is mainly linked to star formation. Survey observations are therefore likely sufficient to identify the non-central sources most likely to heat the ICM.

\end{enumerate}

Our study covers only a relatively limited set of rich clusters, for which high quality multi-wavelength data were available. Our finding that currently available radio surveys capture most of the radio galaxies capable of making significant contributions to ICM heating means that expansion of the sample should not be limited by availability of deep radio observations. More recent radio surveys, e.g., the Very Large Array Sky Survey (VLASS) or the LOFAR Two-metre Sky Survey (LoTSS) may also provide improved resolution and/or sensitivity. Selecting clusters in the SDSS footprint would provide some of the basic optical data needed to identify cluster members. Measurements of ICM pressure may be more problematic, but the archival coverage of the \chandra\ and \textit{XMM-Newton} missions may be sufficient. An expansion of the sample, to gain a more complete picture of the role of non-central radio galaxies in heating the ICM, therefore seems possible.

\section*{Acknowledgements}
This paper is dedicated to the memory of our friend and colleague RS, who carried out the analysis and was preparing to submit the manuscript at the time of her death. On her behalf, the authors thank the University Grants Commission, India, for financial support.

We thank the anonymous referee for their detailed and helpful comments, which have materially improved the paper. EO'S acknowledges the support for this work provided by the National Aeronautics and Space Administration (NASA) through \textit{XMM-Newton} Award Number 80NSSC19K1056. GS acknowledges support for this work provided by the National Aeronautics and Space Administration through Chandra Awards Number GO5-16126X and AR9-20013X issued by the Chandra X-ray Center, which is operated by the Smithsonian Astrophysical Observatory for and on behalf of the National Aeronautics Space Administration under contract NAS8-03060. CPH acknowledges support from ANID through Fondecyt Regular 2021 project no.~1211909. GMRT is a national facility operated by the National Center  for Radio  Astrophysics (NCRA) of the Tata Institute for Fundamental Research (TIFR). This research used SIMBAD and Vizier astronomical database service operated at CDS, Strasbourg. This publication used data from the Two Micron All Sky Survey, jointly operated by University of Massachusetts and the Infrared Processing and Analysis Centre/California Institute of Technology and NASA/IPAC Extragalactic Database (NED) operated by the Jet Propulsion Laboratory, funded by National Aeronautics and Space  Administration and National Science Foundation. This publication also makes use of NASA's Astrophysics Data System Bibliographic Services. This research work made use of the various publicly available Python packages including Numpy, Astropy, and Scipy. TOPCAT is developed by University of Bristol and the erstwhile STARLINK.

\section*{Data Availability}
Raw archival data for GMRT and \chandra{} can be downloaded from \url{https://naps.ncra.tifr.res.in/goa/data/search} and \url{https://cda.harvard.edu/chaser/} respectively. Optical spectroscopic data on galaxies is taken from ACReS and HeCS cluster survey. Data from the ACReS survey including complementary data from other wavelengths are available on request. Data from the HeCS survey is available at \url{https://vizier.u-strasbg.fr/viz-bin/VizieR-3?-source=J/ApJ/767/15/table2}.
Derived radio data and optical properties of galaxies detected in radio are available within the article or its supplementary materials.


\bibliographystyle{mnras}
\bibliography{paper}

\clearpage


\appendix

\section{Notes on the cluster systems}
\label{appendix:cluster notes}

\subsection*{Abell\,383}
Abell\,383, at a redshift of 0.188, is part of the Cluster Lensing And Supernova survey with Hubble \citep[CLASH,][]{Postmanetal12}, and is a dynamically relaxed cluster characterised by a sharp central surface brightness peak and concentric isophotes. An X-ray cavity has also been reported in this cluster \citep{Shinetal16} associated with the central radio source. A combined case study using the data from SZ, X-ray and lensing surveys reveal that this cluster is not spherically symmetric but is extended along the line of sight. \citet{Yuetal18} report that the BCG is well resolved in the 1.5 GHz continuum JVLA image, but concurs with \citet{Hoganetal15a} in finding a steep spectrum diffuse `non-core' radio component, which may suggest the presence of a mini-halo in the cluster core.

\subsection*{Abell\,1758N}
Abell\,1758N, at a redshift of 0.279 is one of two pairs of merging clusters separated by a distance of $\sim$2 Mpc \citep[e.g.][]{Schellenbergeretal19, Monteiro-Oliveiraetal17, DavidKempner04}. Abell\,1758N consists of two components, the NW subcluster being the more massive. \citet{Schellenbergeretal19} detected a shock-front in the NW subcluster implying a merger velocity of a few thousand kms$^{-1}$ and a timescale of $\sim$300-400 Myr post core passage.
This is consistent with the simulations of \citet{Machadoetal15} who predict the two components are currently moving apart and are close to their turnaround point. Several authors including \citet{Schellenbergeretal19} have reported the presence of a giant radio-halo elongated along the merger axis.

\subsection*{Abell\,1914}
Abell\,1914 at a redshift of 0.171 is a well known merging cluster \cite[e.g.][]{Barrenaetal13}. It contains three prominent radio sources, namely a radio halo, a radio Phoenix and a head-tail galaxy \citep{Mandaletal19}. \citet[][]{Botteonetal18} have observed a counter-shock in the cluster and a double tail X-ray morphology much like the bullet cluster. Our X-ray images show a disturbed morphology and asymmetric gas distribution. There are four prominent optical members of the cluster, two of which are the BCGs (neither of which are detected in the radio). The third hosts the bright steep-spectrum \citep{Rolandetal85} radio source 4C38.39 (likely the origin of the plasma now forming the radio Phoenix), and the fourth is a head-tail radio galaxy. Despite the ongoing merger, our 3D pressure profile is smooth and well described by the universal pressure profile.

\subsection*{RXJ\,1720}
RXJ~1720.2+3536 at a redshift of 0.164 \citep{Owersetal11} is a cool-core X-ray luminous galaxy cluster. A beta model subtraction shows a residual in the surface brightness profile suggestive of sloshing motions, probably indicating a recent minor merger. The associated cold-fronts \citep{MazzottaGiacintucci08} are correlated with its prominent radio mini-halo \citep{Giacintuccietal14a}. The central galaxy of the cluster remains unresolved down to sub-kpc scales in the radio.

\subsection*{RXJ\,2129}
RXJ~2129.6+0005 is a CLASH cluster at a redshift of 0.235. It is a relaxed cool-core cluster \citep[e.g.,][]{Tremblayetal15} which hosts a radio mini-halo at its centre \citep{Giacintuccietal19,Kaleetal15b}. The low surface brightness diffuse emission from the mini-halo makes it extremely difficult to clearly detect the BCG at low radio frequencies. The source is point-like at 610 MHz, 1.3 GHz and 4.86 GHz but is resolved in 8.46 GHz VLA observations. \citet{Yuetal18} also note the presence of a `non-core' component in the central source in addition to radio emission generated from the nuclear activity. A faint asymmetric jet extending a few 4 kpc from the bright core component was reported at 8.46~GHz by \citet{Giacintuccietal19} with the possibility of a counter-jet on the other end of the core.

\subsection*{ZwCl\,3146}
ZwCl\,3146 at a redshift of 0.29 is a relaxed  massive cluster, the hottest system in our sample, with a very high cooling rate \citep[$\sim$740\M~yr$^{-1}$,][]{McDonaldetal18}. It is one of the two clusters in our sample with associated X-ray cavities \citep{Raffertyetal06, Shinetal16}. Three prominent cold-fronts along with a mini-halo have been observed \citep{Giacintuccietal14b,Kaleetal15b}. A central radio point source is embedded within the mini-halo, with no indication of radio jets or lobes. \citet{Giacintuccietal14b} show that the radio emission associated with the BCG remains unresolved up to 8.5 GHz in VLA imaging.

\section{Images}
\label{appendix:images}
In this appendix, we present SDSS r-band images  of our clusters, along with images of the 610~MHz continuum emission from the non-central radio sources and 0.5-4 keV X-ray maps. In each image, the left panel shows the SDSS optical image with redshift-confirmed cluster members from ACReS or HeCS marked by blue circles. GMRT 610 MHz contours are overlaid (in pink), spaced by a factor of two, starting from 5$\sigma$, where $\sigma$ is between 50-500 $\mu$Jy beam$^{-1}$. Red circles represent the R$_{200}$ fiducial radius of the cluster. The upper right panels present GMRT 610 MHz contours (cyan) overlaid on the adaptively smoothed, exposure corrected and point source removed \chandra\ ACIS 0.5-4~keV image. The lower right panels show a zoomed in view of the image on the left with \chandra{} ACIS 0.5-4 keV contours (pink) overlaid. Scales are indicated in the captions.

\begin{figure*}
\centering{
\includegraphics[width=0.95\textwidth]{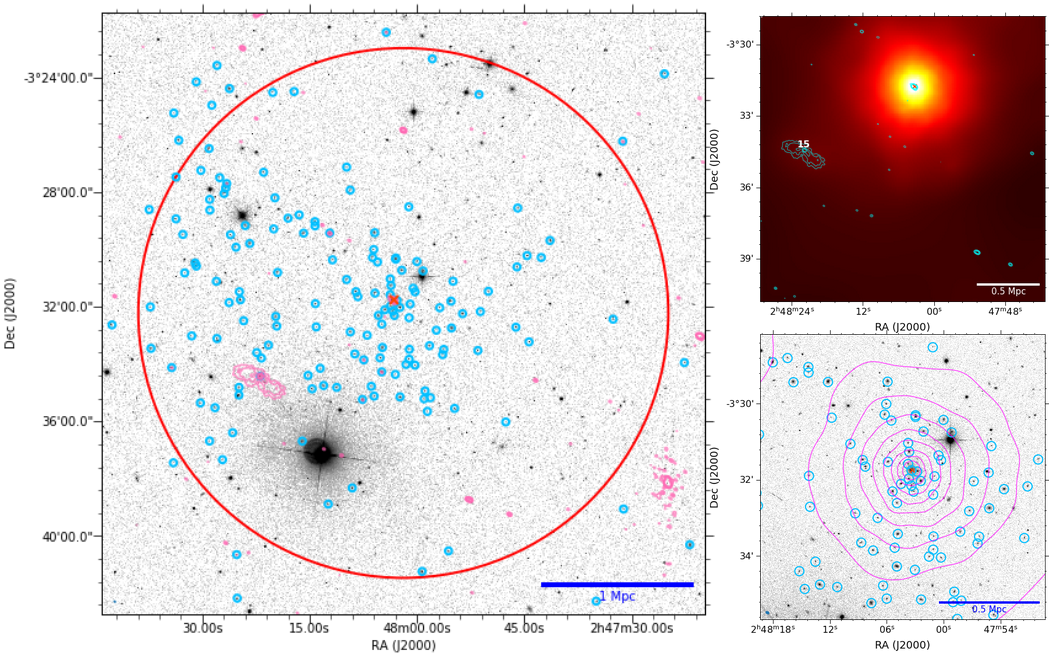}
}
\vspace{-3mm}
\caption{Abell\,383: 1$\sigma$ = 63 $\mu$Jy beam$^{-1}$. Scale used is 3.155 kpc arcsec$^{-1}$.}
\label{fig:a383}
\end{figure*}

\begin{figure*}
\centering{
\includegraphics[width=0.97\textwidth]{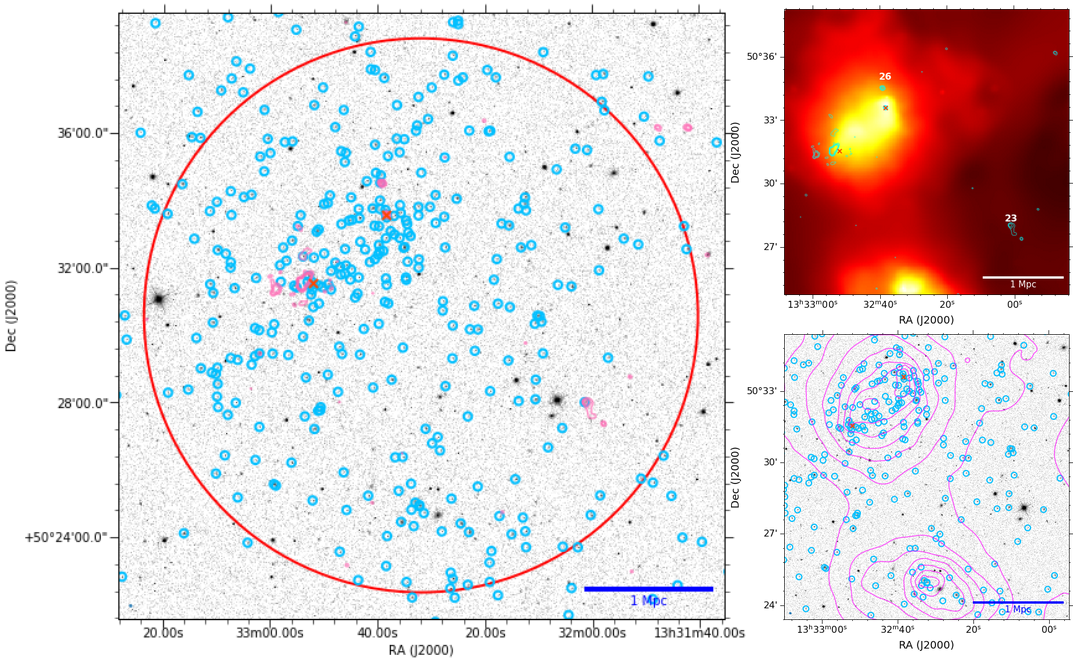}
}
\vspace{-3mm}
\caption{Abell\,1758: 1$\sigma$ = 441 $\mu$Jy beam$^{-1}$. Scale used is 4.234 kpc arcsec$^{-1}$.}
\label{fig:a1758}
\end{figure*}

\begin{figure*}
\centering{
\includegraphics[width=0.97\textwidth]{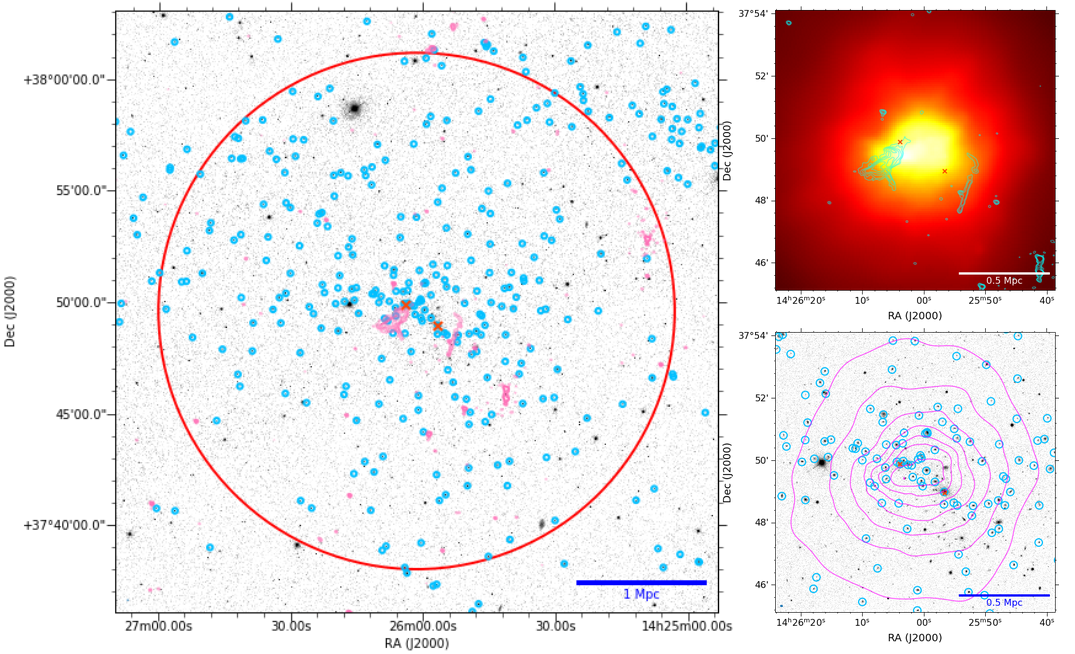}
}
\vspace{-3mm}
\caption{Abell\,1914: 1$\sigma$ = 53 $\mu$Jy beam$^{-1}$. Scale used is 2.857 kpc arcsec$^{-1}$.}
\label{fig:a1914}
\end{figure*}

\begin{figure*}
\centering{
\includegraphics[width=0.97\textwidth]{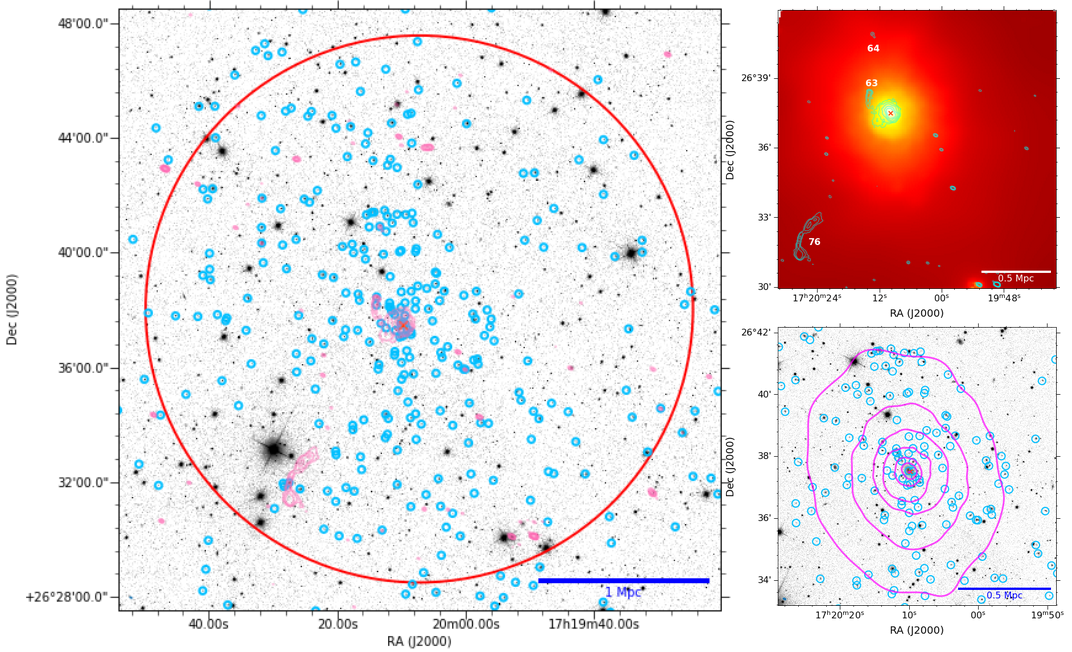}
}
\vspace{-3mm}
\caption{RXJ\,1720: 1$\sigma$ = 69 $\mu$Jy beam$^{-1}$. Scale used is 2.72 kpc arcsec$^{-1}$.}
\label{fig:rxj1720}
\end{figure*}

\begin{figure*}
\centering{
\includegraphics[width=0.95\textwidth]{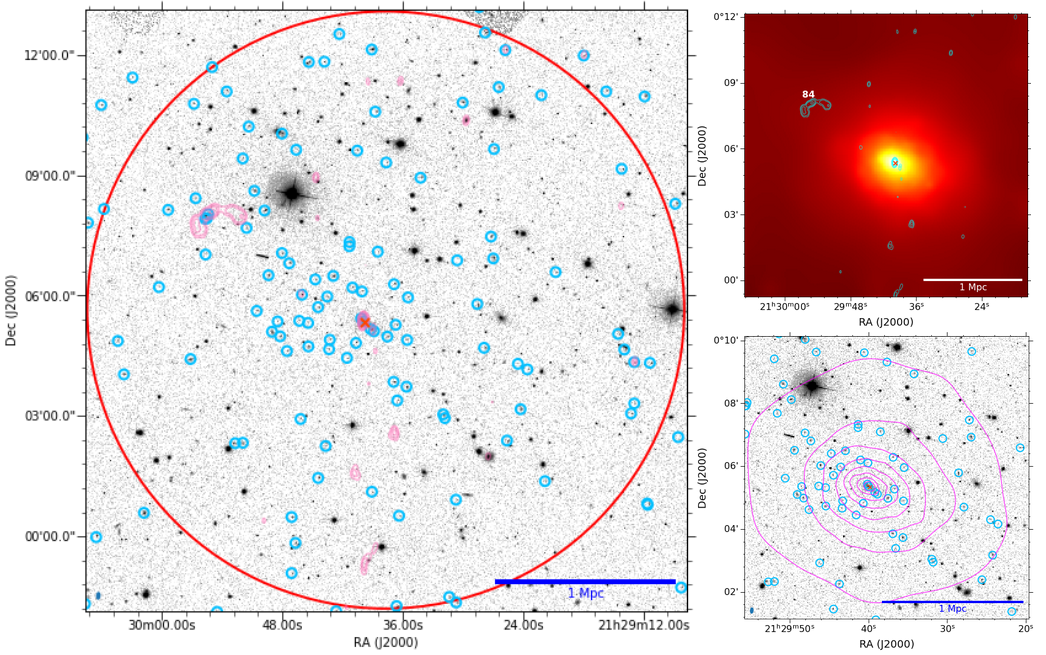}
}
\vspace{-3mm}
\caption{RXJ\,2129: 1$\sigma$ = 450 $\mu$Jy beam$^{-1}$. Scale used is 3.713 kpc arcsec$^{-1}$.}
\label{fig:rxj2129}
\end{figure*}

\begin{figure*}
\centering{
\includegraphics[width=0.98\textwidth]{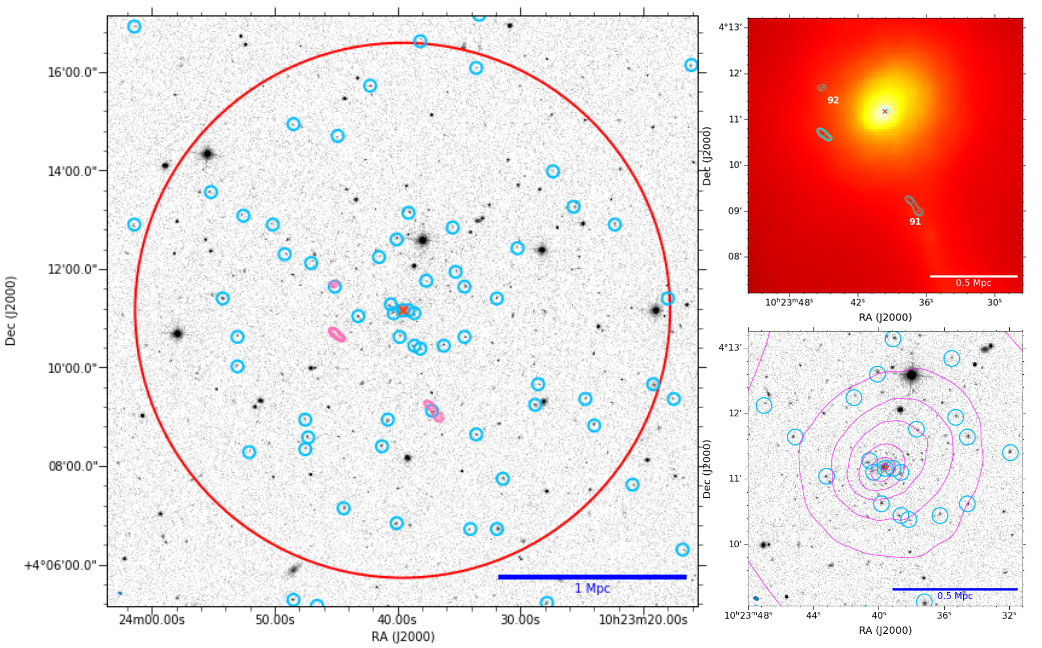}
}
\vspace{-3mm}
\caption{ZwCl\,3146: 1$\sigma$ = 443 $\mu$Jy beam$^{-1}$. Scale used is 4.353 kpc arcsec$^{-1}$.}
\label{fig:z3146}
\end{figure*} 

\clearpage
\onecolumn
\section{Full Table of radio galaxies present in the clusters}
\label{appendix:table}
\input{tables/satellite_populations}

\clearpage
\section{Radio emission expected from the SFR of galaxies}
\input{tables/tableD1}

\bsp	
\label{lastpage}

\twocolumn

\end{document}

%% file: tables/table1.tex
\begin{table*}
    \caption{Galaxy clusters used in this paper. The parameters ICM temperature (T$_x$), cooling luminosity (L$_{\rm{cool}}$) and total X-ray luminosity (L$_x$) are drawn from \citet{Cavagnoloetal09}. Optical properties R$_{200}$, N$_{200}$ and R$_{L_{*}}$ are from the WHL catalogue. M$_{200}$ has been calculated using Eq.2 of \citet{Wenetal12}.
    }
    \label{tab:cluster_table}
	\begin{tabular}{lcccccccccl}
    	\hline
		Cluster Name & RA & Dec & Velocity & T$_{x}$ & L$_{\rm{cool}}$ & L$_{x}$ & R$_{200}$ & R$_{L_{*}}$ & N$_{200}$ & M$_{200}$\\
 		& (hms) & ($\deg$ ,  ', '') & (km~s$^{-1}$) & (keV) &(10$^{42}$erg~s$^{-1}$)  & ($10^{44}\rm{erg~s^{-1}}$) &  (Mpc) & & & (10$^{14}$M$_{\odot}$)\\
    	\hline
		Abell 383 & 02 48 06 & -03 31 45 & 56091 & 3.93 & 4.98 & 5.93 & 1.75& 99.54& 71& 7.04\\

		Abell 1758 & 13 32 34 & 50 30 31 & 83642 & 7.95 & 2.66 & 21.1 & 2.09 & 185.14 & 139 & 14.55\\

		Abell 1914 & 14 26 00 & 37 49 41 & 51324 & 10.5 & 5.11 & 26.4 & 1.99 &139.01 & 118& 10.41\\

		RXJ 1720 & 17 20 10 & 26 40 33 & 49166 & 5.55 & 8.36 & 11.5 & 1.55 & 79.25& 64& 5.39\\

		RXJ 2129 & 21 29 40 & 00 05 21 & 70451 & 6.1 & 6.28 & 12.7 &1.66 &86.84 &62 & 6.00\\

		ZwCl 3146 & 10 23 40 & 04 11 12 & 84092 & 12.8 & 24.83 & 29.8 & 1.42 & 61.42 & 48 & 4.00\\
        \hline
	\end{tabular}
\end{table*}

%% file: tables/table3.tex
\begin{table*}
\begin{minipage}{\linewidth}
 \caption{Details of our GMRT 610~MHz observations analysed for each cluster source. The columns give the cluster name, observation ID, observation date, frequency, bandwidth, time on source, beam parameters and the rms noise in the resulting images.}
\centering
\label{table:GMRTtable}
\begin{tabular}{lccccccc}
 \hline
  Cluster Name & Observation & Observation & Frequency & Bandwidth & On source& Beam,P.A.     & rms \\
          &  ID &  Date    &  (MHz)    &  (MHz)    &  time (hours)    & (Full array, $''\times'',{}^{\circ}$) & mJy beam$^{-1}$ \\
 \hline
  Abell 383  & 33\_034  & 2017 October &   610     &  31.0  &   5.49    &  $6.46\times4.82$, 63.16    &  0.063\\
  Abell 1758  & 23\_023  & 2013 March   &    610    &  32.0  &   5.93   &             $6.46\times5.56$, 69.26   & 0.441 \\
  Abell 1914   & 31\_049 & 2016 December    &    610    &  32.0   &   5.72   &           $5.78\times4.28$, 82.19   & 0.053 \\
  RXJ 1720   & 20\_016 & 2011 July      &    610    &   32.0          & 5.76     &  $5.71\times4.76$, -70.53   & 0.069 \\
   RXJ 2129   & 22\_029 & 2012 September       &    610    &     31.0        & 5.99  &  $12.11\times6.93$, -4.34   & 0.155 \\
  ZwCl 3146   & 19\_039 & 2011 February      &    610    & 32.0 &  5.15   &            $5.80\times3.89$, 65.85    & 0.443 \\
  \hline
\end{tabular}
\end{minipage}
\end{table*}

%% file: tables/X-ray_observations.tex
\begin{table}
\caption{Summary of X-ray observations. All observations were performed with the \chandra\ ACIS-I array, using either the Faint (F) or Very Faint (VF) telemetry modes. Total exposures and exposure times after flare cleaning are listed in a net/gross format. }
\label{table:X-ray_obs}
\begin{center}
\begin{tabular}{lcccc}
\hline
Cluster & Obs. Date & ObsID & Mode & \multicolumn{1}{c}{Exposure} \\
  & & & & (ks) \\
\hline
Abell 383  & 2000-09-08 &   524 & VF & 6.12/9.96\\
           & 2000-11-16 &  2320 & VF & 13.55/19.29\\
Abell 1758  & 2012-09-27 & 13997 & VF & 22.52/27.64 \\
            & 2012-09-28 & 15538 & VF & 75.94/93.33 \\ 
            & 2012-10-09 & 15540 & VF & 20.98/26.72\\
Abell 1914 & 2017-04-29 & 18252 & VF & 24.87/27.69\\
           & 2017-03-09 & 20023 & VF & 23.28/25.42\\
           & 2017-03-11 & 20024 & VF & 15.60/17.17\\
           & 2017-03-06 & 20025 & VF & 19.88/21.68\\
           & 2017-06-02 & 20026 & VF & 27.33/28.87\\
RXJ 1720 & 2002-10-03 & 3224 & VF &  17.16/23.82 \\
         & 2002-08-19 & 4361 & VF &  17.65/25.67 \\
RXJ 2129   & 2000-10-21 &   552 & VF & 6.89/9.96\\
           & 2009-04-03 &  9370 & VF & 25.55/29.64\\
ZwCl 3146     & 2000-05-10 &   909 &  F & 33.76/40.16\\
           & 2008-01-18 &  9371 & VF & 32.51/46.01\\
\hline
\end{tabular}
\end{center}
\end{table}

%% file: tables/xray_properties_galaxies.tex
\renewcommand\tablename{Chart}
\begin{table*}
\caption{The derived quantities estimated for each extended and point-like radio galaxy in our sample. For each radio source, pressure is derived using the universal pressure profile (see \S\ref{subsection:pressure estimation}) the volume is estimated from the radio images (see \S\ref{subsection:volume estimation}) and the sonic expansion timescale (t$_{\rm{sonic}}$) is used as the estimate of source age. P$_{cav}$ is calculated from Equation~\ref{equation:PcavX}, R is the projected cluster-centric radius, r$_{\rm{cool}}$ is the cooling radius and L$_{\rm{cool}}$ is the cooling luminosity.The Galaxy ID number can be used to identify the galaxies in Tables~\ref{tab:C1} and \ref{tab:D1_SFR} and in most cases in the images in Appendix~\ref{appendix:images}.}
\label{tab:x-raytable}
\begin{center}
\begin{tabular}{lccccccccc}
\hline
Galaxy & Cluster & Pressure & Volume & t$_{\rm{sonic}}$ & P$_{cav}$ & R & r$_{\rm{cool}}$ & L$_{\rm{cool}}$ & Radio\\
 ID  &   & (10$^{-11}$~erg~cm$^{-3}$) & (10$^{70}$~cm$^{3}$) & (Myr) & (10$^{42}$~erg~s$^{-1}$) & Mpc & kpc & 10$^{44}$~erg~s$^{-1}$ & morphology\\
\hline
15 & Abell 383 & 3.11 & 11.4 & 164 & 224 & 1.05 & 87 & 4.98 & extended\\
23 & Abell 1758  &  0.02 & 2.17 & 19 & 31 & 2.17$^*$ & 94$^\dag$ & 1.34 & extended\\
26 & Abell 1758 &  3.87 & 0.33 & 8 & 196 & 0.29$^*$ & 94$^\dag$ & 1.34 &point-like\\
33 & Abell 1914 & 0.002 & 0.127 & 29 & 0.101& 4.68 & 121 & 5.11 & extended\\
38 & Abell 1914 & 0.046 & 0.003 & 10 & 0.223& 2.15 & 121 & 5.11 &point-like\\
63 & RXJ 1720  & 10.9 & 1.57 & 82 & 2630 & 0.18 & 122 & 8.36 & extended\\
64 & RXJ 1720  & 2.2 & 0.05 & 20 & 75 & 0.48 & 122 & 8.36 &point-like\\
76 & RXJ 1720 & 0.19 & 4.92 & 171 & 70.3& 1.2 & 122 & 8.36 & extended\\
84 & RXJ 2129  & 0.32 & 22.2 & 110 & 106 & 1.13 & 111 & 6.28 & extended\\
91 & ZwCl 3146  & 2.77 & 0.38 & 41 & 329 & 0.56& 172& 24.8 & extended\\
92 & ZwCl 3146 & 5.81 & 0.12 & 12 & 324 & 0.38 & 172 & 24.8 &point-like\\
\hline
\end{tabular}
\end{center}
* Distance from the NW cluster.
$\dag$ Cooling radius of the NW cluster
\end{table*}

%% file: tables/detection_statistics.tex
\begin{table*}
\caption{Summary of the number and types of radio galaxies detected in each cluster in our sample. The detection statistics gives an estimate of the mean number of radio sources for confirmed optical redshifts expected in the clusters.}
\begin{tabular}{lcccc}
\hspace*{-2cm}\\
\label{tab:detection_statistics}\\
\hline
Cluster & Redshifts measured & Detected in radio & Total fraction & Extended fraction\\
\hline
Abell 383&
266&
22&
0.08&
0.004\\
Abell 1758&
471&
10&
0.02&
0.004\\
Abell 1914&
454&
24&
0.05&
0.004\\
RXJ 1720 &484&24&0.05&0.006\\
RXJ 2129&
334&
10&
0.03&
0.003\\
ZwCl 3146&
98&
5&
0.05&
0.01\\
\hline
\end{tabular}
\end{table*}

%% file: tables/satellite_populations.tex
\begin{ThreePartTable}
\begin{longtable}{lcccccccc}
\caption{\label{tab:C1}Radio flux densities for all detected sources in the GMRT fields of view. The columns list the galaxy ID, RA, Declination, recession velocity, flux density at 610~MHz, radio power at 1.4~GHz, K-band absolute magnitude and radio morphology. All radio sources in our analysis are detected at $>$5 $\times$ r.m.s. Detected central radio sources and non-central sources used in the P$_{\rm cav}$ analysis are marked with $\dagger$ and * respectively.}\\
\hline
\hspace*{-0.1cm}
Galaxy & Cluster & RA & DEC & Recession velocity & S$_{610 MHz}$ & L$_{1.4 GHz}$ & M$_{K}$ & Radio morphology\\
ID  &   &  &  & (kms$^{-1}$) &  $\pm8\%$ (mJy) &  (10$^{23}$ W Hz$^{-1}$) & & \\
\hline
1 & A383 & 2 47 31 & -03 26 14.758 & 58287 & 0.534 & 0.26  & -23.5 & point \\
2 & A383 & 2 47 33 & -03 19 22.895 & 55218 & 0.183 & 0.089 & -24.5 & point \\
3 & A383 & 2 48 04 & -03 22 26.318 & 55899 & 0.417 & 0.203 & -24.1 & point \\
4 & A383 & 2 48 09 & -03 27 56.597 & 55698 & 0.628 & 0.305 & -24.9 & point \\
5 & A383 & 2 48 12 & -03 29 26.882 & 58680 & 1.273 & 0.619 & -25.3 & point \\
6 & A383 & 2 48 27 & -03 27 42.820 & 57684 & 0.366 & 0.178 & -25.1 & point \\
7 & A383 & 2 49 26 & -03 18 15.327 & 56904 & 0.688 & 0.335 & -24.7 & point \\
8 & A383 & 2 46 49 & -03 34 24.113 & 58458 & 0.332 & 0.161 & -24.1 & point \\
9\tnote{$\dagger$} & A383 & 2 48 03 & -03 31 45.807 & 56643 & 76.65 & 37.27 & -26.6 & extended \\
10 & A383 & 2 48 05 & -03 34 17.368 & 54084 & 0.612 & 0.298 & -24.7 & point \\
11 & A383 & 2 47 59 & -03 34 56.593 & 57297 & 0.313 & 0.152 & -23.8 & point \\
12 & A383 & 2 48 25 & -03 31 46.012 & 55653 & 0.277 & 0.135 & -24.8 & point \\
13 & A383 & 2 48 28 & -03 33 07.521 & 57189 & 0.208 & 0.101 & -24.0 & point \\
14 & A383 & 2 48 09 & -03 33 39.508 & 57345 & 0.175 & 0.085 & -23.1 & point \\
15\tnote{*} & A383 & 2 48 22 & -03 34 32.020 & 55899 & 108.5 & 52.77 & -25.2 & extended \\
16 & A383 & 2 48 07 & -03 33 52.099 & 57120 & 0.519 & 0.252 & -23.3 & point \\
17 & A383 & 2 48 34 & -03 34 08.204 & 55731 & 0.469 & 0.228 & -24.1 & point \\
18 & A383 & 2 48 25 & -03 40 39.694 & 56748 & 0.35  & 0.17  & -24.1 & point \\
19 & A383 & 2 48 57 & -03 39 54.912 & 55602 & 0.245 & 0.119 & -23.5 & point \\
20 & A383 & 2 48 08 & -03 35 15.646 & 58140 & 0.479 & 0.233 & -24.4 & point \\
21 & A383 & 2 48 38 & -03 43 58.254 & 57006 & 0.666 & 0.324 & -24.0 & point \\
22 & A383 & 2 49 00 & -03 53 28.410 & 57741 & 0.356 & 0.173 & -24.7 & point \\
23\tnote{*} & A1758 & 13 32 01 & 50 28 0.019  & 83403 & 25.448 & 30.185 & -25.4 & extended \\
24 & A1758 & 13 32 27 & 50 35 17.222 & 82161 & 0.806 & 0.956 & -23.7 & point \\
25\tnote{$\dagger$} & A1758 & 13 32 38 & 50 33 35.310 & 83481 & 17.12 & 20.31 & -26.4 & point \\
26\tnote{*} & A1758 & 13 32 39 & 50 34 31.368 & 88017 & 20.79 & 24.67 & -24.8 & point-like \\
27 & A1758 & 13 32 54 & 50 32 25.135 & 79776 & 12.45 & 14.76 & -24.6 & point \\
28 & A1758 & 13 33 16 & 50 34 31.525 & 87831 & 0.988 & 1.172 & -24.9 & point \\
29 & A1758 & 13 30 33 & 50 31 50.879 & 83703 & 1.043 & 1.238 & -24.5 & point \\
30 & A1758 & 13 30 07 & 50 29 50.102 & 83976 & 4.495 & 5.332 & -25.6 & point \\
31 & A1758 & 13 33 02 & 50 29 27.416 & 83634 & 1.17  & 1.388 & -25.9 & point \\
32 & A1758 & 13 33 32 & 50 22 49.879 & 80121 & 0.819 & 0.971 & -24.1 & point \\
33\tnote{*} & A1914 & 14 27 28 & 38 10 41.051 & 51144 & 9.54  & 3.86  & -25.5 & extended \\
34 & A1914 & 14 24 19 & 37 48 27.788 & 48537 & 0.351 & 0.142 & -24.5 & point \\
35 & A1914 & 14 24 26 & 37 48 01.046 & 48300 & 1.874 & 0.758 & -25.4 & point \\
36 & A1914 & 14 25 09 & 37 51 59.835 & 49005 & 1.109 & 0.449 & -22.5 & point \\
37 & A1914 & 14 25 22 & 37 58 32.580 & 52017 & 0.304 & 0.123 & -24.4 & point \\
38\tnote{*} & A1914 & 14 25 24 & 37 59 38.609 & 49971 & 0.416 & 0.168 & -23.8 & point-like \\
39 & A1914 & 14 25 35 & 37 47 54.413 & 51456 & 0.982 & 0.397 & -24.0 & point \\
40 & A1914 & 14 25 40 & 37 30 39.342 & 50919 & 0.309 & 0.125 & -24.3 & point \\
41 & A1914 & 14 25 45 & 38 01 32.914 & 49422 & 1.396 & 0.565 & -25.7 & point \\
42 & A1914 & 14 25 47 & 37 47 46.715 & 52704 & 3.29  & 1.331 & -23.8 & point \\
43 & A1914 & 14 25 49 & 37 47 38.555 & 50184 & 0.425 & 0.172 & -23.1 & point \\
44 & A1914 & 14 25 49 & 37 42 33.096 & 49425 & 0.533 & 0.216 & -23.8 & point \\
45 & A1914 & 14 25 49 & 37 56 31.364 & 51942 & 0.311 & 0.126 & -24.7 & point \\
46 & A1914 & 14 25 50 & 37 49 55.967 & 50460 & 0.325 & 0.132 & -23.5 & point \\
47 & A1914 & 14 25 57 & 37 37 16.411 & 49110 & 0.227 & 0.092 & -24.2 & point \\
48 & A1914 & 14 25 59 & 37 57 58.713 & 51474 & 0.449 & 0.182 & -24.6 & point \\
49 & A1914 & 14 26 00 & 37 49 21.118 & 49407 & 1.69  & 0.684 & -24.3 & point \\
50 & A1914 & 14 26 01 & 37 45 48.785 & 47805 & 0.491 & 0.199 & -22.9 & point \\
51 & A1914 & 14 26 01 & 37 53 48.015 & 52272 & 0.548 & 0.222 & -23.1 & point \\
52 & A1914 & 14 26 05 & 38 04 06.864 & 49680 & 0.745 & 0.301 & -25.0 & point \\
53 & A1914 & 14 26 11 & 37 50 22.289 & 47991 & 0.249 & 0.101 & -23.7 & point \\
54 & A1914 & 14 26 35 & 37 52 54.437 & 48882 & 0.151 & 0.061 & -23.7 & point \\
55 & A1914 & 14 27 37 & 38 01 25.633 & 49311 & 3.13  & 1.267  & -24.5 & point \\
56 & A1914 & 14 27 59 & 37 52 45.672 & 50343 & 0.357 & 0.144 & -23.2 & point \\
57 & RXJ1720 & 17 20 03 & 26 53 14.619 & 48774 & 2.754 & 1.018 & -25.2 & point \\
58\tnote{$\dagger$} & RXJ1720 & 17 20 10 & 26 37 31.116 & 47991 & 245.577 &  90.73 & -26.6 & extended \\
59 & RXJ1720 & 17 20 07 & 26 37 59.643 & 49095 & 2.305 & 0.852 & -22.2 & point \\
60 & RXJ1720 & 17 19 37 & 26 39 50.955 & 46704 & 0.216 & 0.08 & -23.6 & point \\
61 & RXJ1720 & 17 19 16 & 26 44 28.987 & 46422 & 0.594 & 0.219 & -22.1 & point \\
62 & RXJ1720 & 17 19 15 & 26 44 21.356 & 46746 & 0.276 & 0.102 & -24.4 & point \\
63\tnote{*} & RXJ1720 & 17 20 13 & 26 38 28.039 & 48762 & 17.2 & 6.355 & -25.3 & extended \\
64\tnote{*} & RXJ1720 & 17 20 13 & 26 40 54.449 & 47379 & 2.032 & 0.751 & -24.3 & point-like \\
65 & RXJ1720 & 17 20 32 & 26 40 19.623 & 47658 & 0.827 & 0.306 & -25.8 & point \\
66 & RXJ1720 & 17 20 32 & 26 41 52.948 & 47970 & 0.591 & 0.218 & -25.2 & point \\
67 & RXJ1720 & 17 20 58 & 26 41 26.754 & 48210 & 1.635 & 0.604 & -23.9 & point \\
68 & RXJ1720 & 17 20 00 & 26 35 55.303 & 46944 & 1.045 & 0.386 & -24.4 & point \\
69 & RXJ1720 & 17 19 12 & 26 33 58.333 & 47361 & 0.237 & 0.088 & -23.9 & point \\
70 & RXJ1720 & 17 19 18 & 26 34 01.319 & 47406 & 0.658 & 0.243 & -24.8 & point \\
71 & RXJ1720 & 17 19 26 & 26 33 42.15 & 46140 & 0.298 & 0.11 & -25.9 & point \\
72 & RXJ1720 & 17 19 30 & 26 34 33.004 & 47661 & 1.108 & 0.409 & -24.4 & point \\
73 & RXJ1720 & 17 20 12 & 26 34 20.768 & 49152 & 0.286 & 0.106 & -25.5 & point \\
74 & RXJ1720 & 17 20 54 & 26 34 28.178 & 49452 & 0.261 & 0.096 & -23.4 & point \\
75 & RXJ1720 & 17 20 29 & 26 24 46.242 & 48477 & 0.506 & 0.187 & -23.1 & point \\
76\tnote{*} & RXJ1720 & 17 20 27 & 26 31 57.980 & 47823 & 75.295 & 27.82 & -25.4 & extended \\
77 & RXJ1720 & 17 20 23 & 26 27 28.879 & 48966 & 0.59 & 0.218 & -24.4 & point \\
78 & RXJ1720 & 17 20 13 & 26 30 18.401 & 48795 & 0.463 & 0.171 & -22.8 & point \\
79 & RXJ1720 & 17 21 24 & 26 36 51.955 & 46740 & 0.321 & 0.119 & -24.3 & point \\
80 & RXJ1720 & 17 19 39 & 26 23 12.000 & 48306 & 0.514 & 0.19 & -23.1 & point \\
81 & RXJ2129 & 21 29 26 & 00 12 07.732 & 72246 & 1.14 & 0.92 & -24.4 & point \\
82 & RXJ2129 & 21 29 18 & 00 11 59.936 & 71337 & 1.5 & 1.21 & -24.9 & point \\
83 & RXJ2129 & 21 29 46 & 00 06 03.506 & 71358 & 1.5 & 1.21 & -24.9 & point \\
84\tnote{*} & RXJ2129 & 21 29 56 & 00 08 00.064 & 70209 & 52.177 & 42.102 & -25.7 & extended \\
85 & RXJ2129 & 21 29 55 & 00 11 43.171 & 68259 & 0.491 & 0.396 & -25.1 & point \\
86 & RXJ2129 & 21 28 24 & 00 03 44.310 & 70509 & 2.02 & 1.63 & -25.3 & point \\
87\tnote{$\dagger$} & RXJ2129 & 21 29 40 & 00 05 22.394 & 70167 & 43.2 & 34.858 & -26.9 & point \\
88 & RXJ2129 & 21 29 39 & 00 05 10.101 & 70368 & 2.41 & 1.945 & -25.7 & point \\
89 & RXJ2129 & 21 29 13 & 00 03 21.228 & 70749 & 0.322 & 0.26 & -23.9 & point \\
90 & RXJ2129 & 21 29 13 & 00 04 21.581 & 70614 & 1.75 & 1.412 & -24.7 & point \\
91\tnote{*} & ZwCl 3146 & 10 23 37 & 04 09 05.868 & 86597 & 74.415 & 97.085 & -25.7 & extended \\
92\tnote{*} & ZwCl 3146 & 10 23 45 & 04 11 38.579 & 85603 & 18.827 & 24.562 & -24.4 & point-like \\
93 & ZwCl 3146 & 10 23 48 & 04 14 55.212 & 88479 & 0.38 & 0.496 & -24.4 & point \\
94\tnote{$\dagger$} & ZwCl 3146 & 10 23 39 & 04 11 10.66 & 88342 & 19.1 & 24.918 & -24.7 & point\\
95 & ZwCl 3146 & 10 24 00 & 04 02 15.360 & 87344 & 0.43 & 0.561 & -24.4 & point \\
\hline
\end{longtable}
\begin{tablenotes}\footnotesize
\item[$\dagger$] Central radio source
\item[*] Non-central galaxy used in P$_{\rm cav}$ analysis
\end{tablenotes}
\end{ThreePartTable}

%% file: tables/tableD1.tex
\renewcommand\tablename{Chart}
\begin{ThreePartTable}
\begin{longtable}{lccccc}
 \caption{The table provides star formation rates (SFRs) for all galaxies from 24$\mu$m MIPS observation and the resulting radio power output at 1.4~GHz entirely due to star formation for extended, point and point-like radio sources in our sample computed from 24$\mu$m flux using \citet{Hainesetal11}. The columns provide the galaxy ID, cluster membership, SFR$_{24\mu m}$, expected radio power output at 1.4~GHz, L$_{SFR}$, actual radio power output at 1.4~GHz, L$_{1.4GHz}$, and the radio morphology of the source.} 
 \label{tab:D1_SFR}\\
\hline 
Galaxy &  Cluster &  SFR$_{24\mu m}$  &  L$_{SFR}$ & L$_{1.4GHz}$$^a$ & Radio morphology  \\
ID   &      & (M$_{\odot}$ yr$^{-1}$) &  (10$^{21}$ W Hz$^{-1}$)&   (10$^{21}$ W Hz$^{-1}$) &  \\ 
\hline
1 & A383 & 2.494 & 8.58 & 25.965 & point \\
2 & A383 & 0.0 & 0.0 & 8.898 & point \\
3 & A383 & 2.118 & 7.29 & 20.276 & point \\
4 & A383 & 3.282 & 11.29 & 30.535 & point \\
5 & A383 & 7.521 & 25.87 & 61.899 & point \\
6 & A383 & 4.415 & 15.19 & 17.796 & point \\
7 & A383 & 0.0 & 0.0 & 33.453 & point \\
8 & A383 & 0.0 & 0.0 & 16.143 & point \\
9\tnote{$\dagger$} & A383 & 1.863 & 6.41 & 3726.9 & extended \\
10 & A383 & 2.564 & 8.82 & 29.757 & point \\
11 & A383 & 5.862 & 20.17 & 15.219 & point \\
12 & A383 & 1.117 & 3.84 & 13.469 & point \\
13 & A383 & 6.146 & 21.14 & 10.114 & point \\
14 & A383 & 0.917 & 3.15 & 8.509 & point \\
15 & A383 & 1.125 & 3.87 & 5277.8 & extended \\
16 & A383 & 2.97 & 10.22 & 25.235 & point \\
17 & A383 & 4.731 & 16.27 & 22.804 & point \\
18 & A383 & 21.039 & 72.37 & 17.018 & point \\
19 & A383 & 3.365 & 11.58 & 11.913 & point \\
20 & A383 & 6.314 & 21.72 & 23.29 & point \\
21 & A383 & 5.944 & 20.45 & 32.383 & point \\
22 & A383 & 0.0 & 0.0 & 17.31 & point \\
23 & A1758 & 1.976 & 6.8 & 3018.5 & extended \\
24 & A1758 & 5.091 & 17.51 & 95.606 & point \\
25\tnote{$\dagger$} & A1758 & 0.0 & 0.0 & 2030.7 & point \\
26 & A1758 & 0.0 & 0.0 & 2466.8 & point-like \\
27 & A1758 & 6.721 & 23.12 & 1476.2 & point \\
28 & A1758 & 0.0 & 0.0 & 117.194 & point \\
29 & A1758 & 0.0 & 0.0 & 123.775 & point \\
30 & A1758 & 0.0 & 0.0 & 533.199 & point \\
31 & A1758 & 0.0 & 0.0 & 138.761 & point \\
32 & A1758 & 20.237 & 69.62 & 97.148 & point \\
33 & A1914 & 0.0 & 0.0 & 386.036 & extended \\
34 & A1914 & 0.0 & 0.0 & 14.203 & point \\
35 & A1914 & 0.0 & 0.0 & 75.845 & point \\
36 & A1914 & 0.0 & 0.0 & 44.879 & point \\
37 & A1914 & 0.0 & 0.0 & 12.301 & point \\
38 & A1914 & 4.612 & 15.87 & 16.833 & point-like \\
39 & A1914 & 5.797 & 19.94 & 39.737 & point \\
40 & A1914 & 0.0 & 0.0 & 12.504 & point \\
41 & A1914 & 0.0 & 0.0 & 56.481 & point \\
42 & A1914 & 0.0 & 0.0 & 133.13 & point \\
43 & A1914 & 2.034 & 7.0 & 17.198 & point \\
44 & A1914 & 4.99 & 17.17 & 21.568 & point \\
45 & A1914 & 3.034 & 10.44 & 12.585 & point \\
46 & A1914 & 3.889 & 13.38 & 13.151 & point \\
47 & A1914 & 4.282 & 14.73 & 9.186 & point \\
48 & A1914 & 2.238 & 7.7 & 18.169 & point \\
49 & A1914 & 0.0 & 0.0 & 68.386 & point \\
50 & A1914 & 2.016 & 6.94 & 19.868 & point \\
51 & A1914 & 4.421 & 15.21 & 22.175 & point \\
52 & A1914 & 10.596 & 36.45 & 30.146 & point \\
53 & A1914 & 1.877 & 6.46 & 10.076 & point \\
54 & A1914 & 1.992 & 6.85 & 6.11 & point \\
55 & A1914 & 0.0 & 0.0 & 126.655 & point \\
56 & A1914 & 0.0 & 0.0 & 14.446 & point \\
57 & RXJ1720 & 0.0 & 0.0 & 101.751 & point \\
58\tnote{$\dagger$} & RXJ1720 & 0.0 & 0.0 & 9073.5 & extended \\
59 & RXJ1720 & 0.0 & 0.0 & 85.18 & point \\
60 & RXJ1720 & 7.207 & 24.79 & 7.981 & point \\
61 & RXJ1720 & 2.853 & 9.81 & 21.947 & point \\
62 & RXJ1720 & 2.635 & 9.06 & 10.198 & point \\
63 & RXJ1720 & 0.0 & 0.0 & 635.497 & extended \\
64 & RXJ1720 & 0.0 & 0.0 & 75.085 & point-like \\
65 & RXJ1720 & 2.671 & 9.19 & 30.556 & point \\
66 & RXJ1720 & 0.0 & 0.0 & 21.836 & point \\
67 & RXJ1720 & 0.0 & 0.0 & 60.416 & point \\
68 & RXJ1720 & 7.96 & 27.38 & 38.606 & point \\
69 & RXJ1720 & 9.942 & 34.2 & 8.757 & point \\
70 & RXJ1720 & 7.487 & 25.76 & 24.311 & point \\
71 & RXJ1720 & 2.52 & 8.67 & 11.01 & point \\
72 & RXJ1720 & 8.908 & 30.64 & 40.937 & point \\
73 & RXJ1720 & 0.0 & 0.0 & 10.567 & point \\
74 & RXJ1720 & 0.936 & 3.22 & 9.643 & point \\
75 & RXJ1720 & 2.561 & 8.81 & 18.695 & point \\
76 & RXJ1720 & 0.0 & 0.0 & 2781.9 & extended \\
77 & RXJ1720 & 6.697 & 23.04 & 21.799 & point \\
78 & RXJ1720 & 1.011 & 3.48 & 17.107 & point \\
79 & RXJ1720 & 0.0 & 0.0 & 11.86 & point \\
80 & RXJ1720 & 0.0 & 0.0 & 18.991 & point \\
81 & RXJ2129 & 2.347 & 8.07 & 91.987 & point \\
82 & RXJ2129 & 26.401 & 90.82 & 121.036 & point \\
83 & RXJ2129 & 0.0 & 0.0 & 121.036 & point \\
84 & RXJ2129 & 0.0 & 0.0 & 4210.2 & extended \\
85 & RXJ2129 & 2.674 & 9.2 & 39.62 & point \\
86 & RXJ2129 & 0.0 & 0.0 & 162.99 & point \\
87\tnote{$\dagger$} & RXJ2129 & 3.144 & 10.82 & 3485.8 & point \\
88 & RXJ2129 & 0.0 & 0.0 & 194.46 & point \\
89 & RXJ2129 & 11.2 & 38.53 & 25.98 & point \\
90 & RXJ2129 & 17.107 & 58.85 & 141.21 & point \\
91 & ZwCl3146 & 0.0 & 0.0 &  9708.5    & extended \\
92 & ZwCl3146 & 0.0 & 0.0 & 2456.2     & point-like \\
93 & ZwCl3146 & 0.0 & 0.0 & 49.6     & point \\
94\tnote{$\dagger$} & ZwCl3146 & 0.0 & 0.0 & 2491.8      & point \\
95 & ZwCl3146 & 0.0 & 0.0 & 56.1     & point \\ \hline
\end{longtable}
\begin{tablenotes}\footnotesize
\item[$^a$] L$_{1.4GHz}$ was calculated here by extrapolating the 1.4~GHz flux density from the available 610~MHz emission, using a spectral index of -0.8
\item[$\dagger$] Central radio source
\end{tablenotes}
\end{ThreePartTable}

%% file: ruchika_paper.bbl
\begin{thebibliography}{}
\makeatletter
\relax
\def\mn@urlcharsother{\let\do\@makeother \do\$\do\&\do\#\do\^\do\_\do\%\do\~}
\def\mn@doi{\begingroup\mn@urlcharsother \@ifnextchar [ {\mn@doi@}
  {\mn@doi@[]}}
\def\mn@doi@[#1]#2{\def\@tempa{#1}\ifx\@tempa\@empty \href
  {http://dx.doi.org/#2} {doi:#2}\else \href {http://dx.doi.org/#2} {#1}\fi
  \endgroup}
\def\mn@eprint#1#2{\mn@eprint@#1:#2::\@nil}
\def\mn@eprint@arXiv#1{\href {http://arxiv.org/abs/#1} {{\tt arXiv:#1}}}
\def\mn@eprint@dblp#1{\href {http://dblp.uni-trier.de/rec/bibtex/#1.xml}
  {dblp:#1}}
\def\mn@eprint@#1:#2:#3:#4\@nil{\def\@tempa {#1}\def\@tempb {#2}\def\@tempc
  {#3}\ifx \@tempc \@empty \let \@tempc \@tempb \let \@tempb \@tempa \fi \ifx
  \@tempb \@empty \def\@tempb {arXiv}\fi \@ifundefined
  {mn@eprint@\@tempb}{\@tempb:\@tempc}{\expandafter \expandafter \csname
  mn@eprint@\@tempb\endcsname \expandafter{\@tempc}}}

\bibitem[\protect\citeauthoryear{{Arnaud}, {Pratt}, {Piffaretti},
  {B{\"o}hringer}, {Croston}  \& {Pointecouteau}}{{Arnaud}
  et~al.}{2010}]{Arnaudetal10}
{Arnaud} M.,  {Pratt} G.~W.,  {Piffaretti} R.,  {B{\"o}hringer} H.,  {Croston}
  J.~H.,   {Pointecouteau} E.,  2010, \mn@doi [A\&A]
  {10.1051/0004-6361/200913416}, \href
  {https://ui.adsabs.harvard.edu/abs/2010A&A...517A..92A} {517, A92}

\bibitem[\protect\citeauthoryear{{Bandara}, {Crampton}  \& {Simard}}{{Bandara}
  et~al.}{2009}]{Bandaraetal09}
{Bandara} K.,  {Crampton} D.,   {Simard} L.,  2009, \mn@doi [ApJ]
  {10.1088/0004-637X/704/2/1135}, \href
  {https://ui.adsabs.harvard.edu/abs/2009ApJ...704.1135B} {704, 1135}

\bibitem[\protect\citeauthoryear{{Barrena}, {Girardi}  \& {Boschin}}{{Barrena}
  et~al.}{2013}]{Barrenaetal13}
{Barrena} R.,  {Girardi} M.,   {Boschin} W.,  2013, \mn@doi [MNRAS]
  {10.1093/mnras/stt144}, \href
  {https://ui.adsabs.harvard.edu/abs/2013MNRAS.430.3453B} {430, 3453}

\bibitem[\protect\citeauthoryear{{Begelman}, {Rees}  \& {Blandford}}{{Begelman}
  et~al.}{1979}]{Begelmanetal79}
{Begelman} M.~C.,  {Rees} M.~J.,   {Blandford} R.~D.,  1979, \mn@doi [Nature]
  {10.1038/279770a0}, \href
  {https://ui.adsabs.harvard.edu/abs/1979Natur.279..770B} {279, 770}

\bibitem[\protect\citeauthoryear{{Bell}}{{Bell}}{2003}]{Belletal03}
{Bell} E.~F.,  2003, \mn@doi [ApJ] {10.1086/367829}, \href
  {http://adsabs.harvard.edu/abs/2003ApJ...586..794B} {586, 794}

\bibitem[\protect\citeauthoryear{{Best} \& {Heckman}}{{Best} \&
  {Heckman}}{2012}]{BestHeckman12}
{Best} P.~N.,  {Heckman} T.~M.,  2012, \mn@doi [MNRAS]
  {10.1111/j.1365-2966.2012.20414.x}, \href
  {https://ui.adsabs.harvard.edu/abs/2012MNRAS.421.1569B} {421, 1569}

\bibitem[\protect\citeauthoryear{{Best}, {Kauffmann}, {Heckman}  \&
  {Ivezi{\'c}}}{{Best} et~al.}{2005}]{Bestetal05}
{Best} P.~N.,  {Kauffmann} G.,  {Heckman} T.~M.,   {Ivezi{\'c}} {\v{Z}}.,
  2005, \mn@doi [MNRAS] {10.1111/j.1365-2966.2005.09283.x}, \href
  {https://ui.adsabs.harvard.edu/abs/2005MNRAS.362....9B} {362, 9}

\bibitem[\protect\citeauthoryear{{Best}, {von der Linden}, {Kauffmann},
  {Heckman}  \& {Kaiser}}{{Best} et~al.}{2007}]{Bestetal07}
{Best} P.~N.,  {von der Linden} A.,  {Kauffmann} G.,  {Heckman} T.~M.,
  {Kaiser} C.~R.,  2007, \mn@doi [MNRAS] {10.1111/j.1365-2966.2007.11937.x},
  \href {http://adsabs.harvard.edu/abs/2007MNRAS.379..894B} {379, 894}

\bibitem[\protect\citeauthoryear{{B{\^i}rzan}, {Rafferty}, {McNamara}, {Wise}
  \& {Nulsen}}{{B{\^i}rzan} et~al.}{2004}]{Birzanetal04}
{B{\^i}rzan} L.,  {Rafferty} D.~A.,  {McNamara} B.~R.,  {Wise} M.~W.,
  {Nulsen} P.~E.~J.,  2004, \mn@doi [ApJ] {10.1086/383519}, \href
  {http://adsabs.harvard.edu/cgi-bin/nph-bib_query?bibcode=2004ApJ...607..800B&db_key=AST}
  {607, 800}

\bibitem[\protect\citeauthoryear{{B{\^i}rzan}, {McNamara}, {Nulsen}, {Carilli}
  \& {Wise}}{{B{\^i}rzan} et~al.}{2008}]{Birzanetal08}
{B{\^i}rzan} L.,  {McNamara} B.~R.,  {Nulsen} P.~E.~J.,  {Carilli} C.~L.,
  {Wise} M.~W.,  2008, \mn@doi [ApJ] {10.1086/591416}, \href
  {http://adsabs.harvard.edu/abs/2008ApJ...686..859B} {686, 859}

\bibitem[\protect\citeauthoryear{{Botteon}, {Gastaldello}  \&
  {Brunetti}}{{Botteon} et~al.}{2018}]{Botteonetal18}
{Botteon} A.,  {Gastaldello} F.,   {Brunetti} G.,  2018, \mn@doi [MNRAS]
  {10.1093/mnras/sty598}, \href
  {https://ui.adsabs.harvard.edu/abs/2018MNRAS.476.5591B} {476, 5591}

\bibitem[\protect\citeauthoryear{{Cavagnolo}, {Donahue}, {Voit}  \&
  {Sun}}{{Cavagnolo} et~al.}{2009}]{Cavagnoloetal09}
{Cavagnolo} K.~W.,  {Donahue} M.,  {Voit} G.~M.,   {Sun} M.,  2009, \mn@doi
  [ApJS] {10.1088/0067-0049/182/1/12}, \href
  {http://adsabs.harvard.edu/abs/2009ApJS..182...12C} {182, 12}

\bibitem[\protect\citeauthoryear{{Clarke} et~al.,}{{Clarke}
  et~al.}{2019}]{Clarkeetal19}
{Clarke} A.~O.,  et~al., 2019, \mn@doi [A\&A] {10.1051/0004-6361/201935584},
  \href {https://ui.adsabs.harvard.edu/abs/2019A&A...627A.176C} {627, A176}

\bibitem[\protect\citeauthoryear{{Condon}}{{Condon}}{1992}]{Condonetal92}
{Condon} J.~J.,  1992, \mn@doi [ARA\&A] {10.1146/annurev.aa.30.090192.003043},
  \href {https://ui.adsabs.harvard.edu/abs/1992ARA&A..30..575C} {30, 575}

\bibitem[\protect\citeauthoryear{{Croston}, {Birkinshaw}, {Hardcastle}  \&
  {Worrall}}{{Croston} et~al.}{2004}]{Crostonetal04}
{Croston} J.~H.,  {Birkinshaw} M.,  {Hardcastle} M.~J.,   {Worrall} D.~M.,
  2004, \mn@doi [MNRAS] {10.1111/j.1365-2966.2004.08118.x}, \href
  {https://ui.adsabs.harvard.edu/abs/2004MNRAS.353..879C} {353, 879}

\bibitem[\protect\citeauthoryear{{Croston}, {Hardcastle}, {Birkinshaw},
  {Worrall}  \& {Laing}}{{Croston} et~al.}{2008}]{Crostonetal08}
{Croston} J.~H.,  {Hardcastle} M.~J.,  {Birkinshaw} M.,  {Worrall} D.~M.,
  {Laing} R.~A.,  2008, \mn@doi [MNRAS] {10.1111/j.1365-2966.2008.13162.x},
  \href {http://adsabs.harvard.edu/abs/2008MNRAS.386.1709C} {386, 1709}

\bibitem[\protect\citeauthoryear{{David} \& {Kempner}}{{David} \&
  {Kempner}}{2004}]{DavidKempner04}
{David} L.~P.,  {Kempner} J.,  2004, \mn@doi [ApJ] {10.1086/423195}, \href
  {https://ui.adsabs.harvard.edu/abs/2004ApJ...613..831D} {613, 831}

\bibitem[\protect\citeauthoryear{{Dong}, {Rasmussen}  \& {Mulchaey}}{{Dong}
  et~al.}{2010}]{Dongetal10}
{Dong} R.,  {Rasmussen} J.,   {Mulchaey} J.~S.,  2010, \mn@doi [ApJ]
  {10.1088/0004-637X/712/2/883}, \href
  {http://adsabs.harvard.edu/abs/2010ApJ...712..883D} {712, 883}

\bibitem[\protect\citeauthoryear{{Dunn}, {Fabian}  \& {Taylor}}{{Dunn}
  et~al.}{2005}]{Dunnetal05}
{Dunn} R.~J.~H.,  {Fabian} A.~C.,   {Taylor} G.~B.,  2005, \mn@doi [MNRAS]
  {10.1111/j.1365-2966.2005.09673.x}, \href
  {http://adsabs.harvard.edu/abs/2005MNRAS.364.1343D} {364, 1343}

\bibitem[\protect\citeauthoryear{{Eckert}, {Gaspari}, {Gastaldello}, {Le Brun}
  \& {O'Sullivan}}{{Eckert} et~al.}{2021}]{Eckertetal21}
{Eckert} D.,  {Gaspari} M.,  {Gastaldello} F.,  {Le Brun} A. M.~C.,
  {O'Sullivan} E.,  2021, \mn@doi [Universe] {10.3390/universe7050142}, \href
  {https://ui.adsabs.harvard.edu/abs/2021Univ....7..142E} {7, 142}

\bibitem[\protect\citeauthoryear{{Fabian}}{{Fabian}}{2012}]{Fabian12}
{Fabian} A.~C.,  2012, ARA\&A, 50, 455

\bibitem[\protect\citeauthoryear{{Fabian}, {Nulsen}  \& {Canizares}}{{Fabian}
  et~al.}{1984}]{Fabianetal84}
{Fabian} A.~C.,  {Nulsen} P.~E.~J.,   {Canizares} C.~R.,  1984, \mn@doi
  [Nature] {10.1038/310733a0}, \href
  {https://ui.adsabs.harvard.edu/abs/1984Natur.310..733F} {310, 733}

\bibitem[\protect\citeauthoryear{{Fanaroff} \& {Riley}}{{Fanaroff} \&
  {Riley}}{1974}]{FanaroffRiley74}
{Fanaroff} B.~L.,  {Riley} J.~M.,  1974, \mn@doi [MNRAS]
  {10.1093/mnras/167.1.31P}, 167, 31

\bibitem[\protect\citeauthoryear{{Ferrarese}}{{Ferrarese}}{2002}]{Ferrarese02}
{Ferrarese} L.,  2002, \mn@doi [ApJ] {10.1086/342308}, \href
  {https://ui.adsabs.harvard.edu/abs/2002ApJ...578...90F} {578, 90}

\bibitem[\protect\citeauthoryear{{Finoguenov} \& {Jones}}{{Finoguenov} \&
  {Jones}}{2001}]{FinoguenovJones01}
{Finoguenov} A.,  {Jones} C.,  2001, \mn@doi [ApJ] {10.1086/318910}, \href
  {https://ui.adsabs.harvard.edu/abs/2001ApJ...547L.107F} {547, L107}

\bibitem[\protect\citeauthoryear{{Forman} et~al.,}{{Forman}
  et~al.}{2005}]{Formanetal05}
{Forman} W.,  et~al., 2005, \mn@doi [ApJ] {10.1086/429746}, \href
  {http://adsabs.harvard.edu/cgi-bin/nph-bib_query?bibcode=2005ApJ...635..894F&db_key=AST}
  {635, 894}

\bibitem[\protect\citeauthoryear{{Gendron-Marsolais}
  et~al.,}{{Gendron-Marsolais} et~al.}{2020}]{GendronMarsolaisetal20}
{Gendron-Marsolais} M.,  et~al., 2020, \mn@doi [MNRAS]
  {10.1093/mnras/staa2003}, \href
  {https://ui.adsabs.harvard.edu/abs/2020MNRAS.499.5791G} {499, 5791}

\bibitem[\protect\citeauthoryear{{Giacintucci}, {Markevitch}, {Venturi},
  {Clarke}, {Cassano}  \& {Mazzotta}}{{Giacintucci}
  et~al.}{2014a}]{Giacintuccietal14b}
{Giacintucci} S.,  {Markevitch} M.,  {Venturi} T.,  {Clarke} T.~E.,  {Cassano}
  R.,   {Mazzotta} P.,  2014a, \mn@doi [ApJ] {10.1088/0004-637X/781/1/9}, \href
  {https://ui.adsabs.harvard.edu/abs/2014ApJ...781....9G} {781, 9}

\bibitem[\protect\citeauthoryear{{Giacintucci}, {Markevitch}, {Brunetti},
  {ZuHone}, {Venturi}, {Mazzotta}  \& {Bourdin}}{{Giacintucci}
  et~al.}{2014b}]{Giacintuccietal14a}
{Giacintucci} S.,  {Markevitch} M.,  {Brunetti} G.,  {ZuHone} J.~A.,  {Venturi}
  T.,  {Mazzotta} P.,   {Bourdin} H.,  2014b, \mn@doi [ApJ]
  {10.1088/0004-637X/795/1/73}, \href
  {https://ui.adsabs.harvard.edu/abs/2014ApJ...795...73G} {795, 73}

\bibitem[\protect\citeauthoryear{{Giacintucci}, {Markevitch}, {Cassano},
  {Venturi}, {Clarke}, {Kale}  \& {Cuciti}}{{Giacintucci}
  et~al.}{2019}]{Giacintuccietal19}
{Giacintucci} S.,  {Markevitch} M.,  {Cassano} R.,  {Venturi} T.,  {Clarke}
  T.~E.,  {Kale} R.,   {Cuciti} V.,  2019, \mn@doi [ApJ]
  {10.3847/1538-4357/ab29f1}, \href
  {https://ui.adsabs.harvard.edu/abs/2019ApJ...880...70G} {880, 70}

\bibitem[\protect\citeauthoryear{{Giacintucci}, {Markevitch},
  {Johnston-Hollitt}, {Wik}, {Wang}  \& {Clarke}}{{Giacintucci}
  et~al.}{2020}]{Giacintuccietal20}
{Giacintucci} S.,  {Markevitch} M.,  {Johnston-Hollitt} M.,  {Wik} D.~R.,
  {Wang} Q.~H.~S.,   {Clarke} T.~E.,  2020, \mn@doi [ApJ]
  {10.3847/1538-4357/ab6a9d}, \href
  {https://ui.adsabs.harvard.edu/abs/2020ApJ...891....1G} {891, 1}

\bibitem[\protect\citeauthoryear{{Gitti}, {Brighenti}  \& {McNamara}}{{Gitti}
  et~al.}{2012}]{Gittietal12}
{Gitti} M.,  {Brighenti} F.,   {McNamara} B.~R.,  2012, \mn@doi [Advances in
  Astronomy] {10.1155/2012/950641}, \href
  {https://ui-adsabs-harvard-edu.ezp-prod1.hul.harvard.edu/abs/2012AdAst2012E...6G}
  {2012, 950641}

\bibitem[\protect\citeauthoryear{{Greisen}}{{Greisen}}{1990}]{Greisen90}
{Greisen} E.~W.,  1990, in Acquisition, Processing and Archiving of
  Astronomical Images. pp 125--142

\bibitem[\protect\citeauthoryear{{Haines}, {Busarello}, {Merluzzi}, {Smith},
  {Raychaudhury}, {Mercurio}  \& {Smith}}{{Haines} et~al.}{2011}]{Hainesetal11}
{Haines} C.~P.,  {Busarello} G.,  {Merluzzi} P.,  {Smith} R.~J.,
  {Raychaudhury} S.,  {Mercurio} A.,   {Smith} G.~P.,  2011, \mn@doi [MNRAS]
  {10.1111/j.1365-2966.2010.17892.x}, \href
  {https://ui.adsabs.harvard.edu/abs/2011MNRAS.412..145H} {412, 145}

\bibitem[\protect\citeauthoryear{{Haines} et~al.,}{{Haines}
  et~al.}{2013}]{Hainesetal13}
{Haines} C.~P.,  et~al., 2013, \mn@doi [ApJ] {10.1088/0004-637X/775/2/126},
  \href {https://ui.adsabs.harvard.edu/abs/2013ApJ...775..126H} {775, 126}

\bibitem[\protect\citeauthoryear{{Hardcastle} et~al.,}{{Hardcastle}
  et~al.}{2019}]{Hardcastleetal19}
{Hardcastle} M.~J.,  et~al., 2019, \mn@doi [A\&A]
  {10.1051/0004-6361/201833893}, 622, A12

\bibitem[\protect\citeauthoryear{{Hickox} et~al.,}{{Hickox}
  et~al.}{2009}]{Hickoxetal09}
{Hickox} R.~C.,  et~al., 2009, \mn@doi [ApJ] {10.1088/0004-637X/696/1/891},
  \href {https://ui.adsabs.harvard.edu/abs/2009ApJ...696..891H} {696, 891}

\bibitem[\protect\citeauthoryear{{Hogan} et~al.,}{{Hogan}
  et~al.}{2015}]{Hoganetal15a}
{Hogan} M.~T.,  et~al., 2015, \mn@doi [MNRAS] {10.1093/mnras/stv1517}, \href
  {https://ui.adsabs.harvard.edu/abs/2015MNRAS.453.1201H} {453, 1201}

\bibitem[\protect\citeauthoryear{{Kale}, {Venturi}, {Cassano}, {Giacintucci},
  {Bardelli}, {Dallacasa}  \& {Zucca}}{{Kale} et~al.}{2015}]{Kaleetal15b}
{Kale} R.,  {Venturi} T.,  {Cassano} R.,  {Giacintucci} S.,  {Bardelli} S.,
  {Dallacasa} D.,   {Zucca} E.,  2015, \mn@doi [A\&A]
  {10.1051/0004-6361/201526341}, \href
  {https://ui.adsabs.harvard.edu/abs/2015A&A...581A..23K} {581, A23}

\bibitem[\protect\citeauthoryear{{Kellermann} \& {Owen}}{{Kellermann} \&
  {Owen}}{1988}]{KellermannOwen98}
{Kellermann} K.~I.,  {Owen} F.~N.,  1988, {Radio galaxies and quasars.}.
Spinger-Verlag, pp 563--602

\bibitem[\protect\citeauthoryear{{Leccardi} \& {Molendi}}{{Leccardi} \&
  {Molendi}}{2008}]{LeccardiMolendi08}
{Leccardi} A.,  {Molendi} S.,  2008, \mn@doi [A\&A]
  {10.1051/0004-6361:200809538}, \href
  {https://ui.adsabs.harvard.edu/abs/2008A&A...486..359L} {486, 359}

\bibitem[\protect\citeauthoryear{{Li}, {Su}  \& {Jones}}{{Li}
  et~al.}{2018}]{Lietal18b}
{Li} Y.,  {Su} Y.,   {Jones} C.,  2018, \mn@doi [MNRAS]
  {10.1093/mnras/sty2125}, \href
  {https://ui.adsabs.harvard.edu/abs/2018MNRAS.480.4279L} {480, 4279}

\bibitem[\protect\citeauthoryear{{Lin} \& {Mohr}}{{Lin} \&
  {Mohr}}{2007}]{LinMohr07}
{Lin} Y.-T.,  {Mohr} J.~J.,  2007, \mn@doi [ApJS] {10.1086/513565}, \href
  {https://ui.adsabs.harvard.edu/abs/2007ApJS..170...71L} {170, 71}

\bibitem[\protect\citeauthoryear{{Liu} et~al.,}{{Liu} et~al.}{2020}]{Liuetal20}
{Liu} W.,  et~al., 2020, \mn@doi [MNRAS] {10.1093/mnras/staa005}, \href
  {https://ui.adsabs.harvard.edu/abs/2020MNRAS.492.3156L} {492, 3156}

\bibitem[\protect\citeauthoryear{{Machacek}, {Nulsen}, {Jones}  \&
  {Forman}}{{Machacek} et~al.}{2006}]{Machaceketal06}
{Machacek} M.,  {Nulsen} P.~E.~J.,  {Jones} C.,   {Forman} W.~R.,  2006,
  \mn@doi [ApJ] {10.1086/505963}, \href
  {http://adsabs.harvard.edu/cgi-bin/nph-bib_query?bibcode=2006ApJ...648..947M&db_key=AST}
  {648, 947}

\bibitem[\protect\citeauthoryear{{Machado}, {Monteiro-Oliveira}, {Lima Neto}
  \& {Cypriano}}{{Machado} et~al.}{2015}]{Machadoetal15}
{Machado} R.~E.~G.,  {Monteiro-Oliveira} R.,  {Lima Neto} G.~B.,   {Cypriano}
  E.~S.,  2015, \mn@doi [MNRAS] {10.1093/mnras/stv1162}, \href
  {https://ui.adsabs.harvard.edu/abs/2015MNRAS.451.3309M} {451, 3309}

\bibitem[\protect\citeauthoryear{{Mandal} et~al.,}{{Mandal}
  et~al.}{2019}]{Mandaletal19}
{Mandal} S.,  et~al., 2019, \mn@doi [A\&A] {10.1051/0004-6361/201833992}, \href
  {https://ui.adsabs.harvard.edu/abs/2019A&A...622A..22M} {622, A22}

\bibitem[\protect\citeauthoryear{{Mantz}, {Allen}, {Morris}, {Rapetti},
  {Applegate}, {Kelly}, {von der Linden}  \& {Schmidt}}{{Mantz}
  et~al.}{2014}]{Mantzetal14}
{Mantz} A.~B.,  {Allen} S.~W.,  {Morris} R.~G.,  {Rapetti} D.~A.,  {Applegate}
  D.~E.,  {Kelly} P.~L.,  {von der Linden} A.,   {Schmidt} R.~W.,  2014,
  \mn@doi [MNRAS] {10.1093/mnras/stu368}, \href
  {https://ui.adsabs.harvard.edu/abs/2014MNRAS.440.2077M} {440, 2077}

\bibitem[\protect\citeauthoryear{{Mazzotta} \& {Giacintucci}}{{Mazzotta} \&
  {Giacintucci}}{2008}]{MazzottaGiacintucci08}
{Mazzotta} P.,  {Giacintucci} S.,  2008, \mn@doi [ApJ] {10.1086/529433}, \href
  {http://adsabs.harvard.edu/abs/2008ApJ...675L...9M} {675, L9}

\bibitem[\protect\citeauthoryear{{McDonald}, {Gaspari}, {McNamara}  \&
  {Tremblay}}{{McDonald} et~al.}{2018}]{McDonaldetal18}
{McDonald} M.,  {Gaspari} M.,  {McNamara} B.~R.,   {Tremblay} G.~R.,  2018,
  \mn@doi [ApJ] {10.3847/1538-4357/aabace}, \href
  {https://ui.adsabs.harvard.edu/abs/2018ApJ...858...45M} {858, 45}

\bibitem[\protect\citeauthoryear{{McNamara} \& {Nulsen}}{{McNamara} \&
  {Nulsen}}{2007}]{McNamaraNulsen07}
{McNamara} B.~R.,  {Nulsen} P.~E.~J.,  2007, \mn@doi [ARA\&A]
  {10.1146/annurev.astro.45.051806.110625}, \href
  {http://adsabs.harvard.edu/abs/2007ARA%26A..45..117M} {45, 117}

\bibitem[\protect\citeauthoryear{{McNamara} \& {Nulsen}}{{McNamara} \&
  {Nulsen}}{2012}]{McNamaraNulsen12}
{McNamara} B.~R.,  {Nulsen} P.~E.~J.,  2012, \mn@doi [New Journal of Physics]
  {10.1088/1367-2630/14/5/055023}, \href
  {http://adsabs.harvard.edu/abs/2012NJPh...14e5023M} {14, 055023}

\bibitem[\protect\citeauthoryear{{McNamara}, {Nulsen}, {Wise}, {Rafferty},
  {Carilli}, {Sarazin}  \& {Blanton}}{{McNamara} et~al.}{2005}]{McNamaraetal05}
{McNamara} B.~R.,  {Nulsen} P.~E.~J.,  {Wise} M.~W.,  {Rafferty} D.~A.,
  {Carilli} C.,  {Sarazin} C.~L.,   {Blanton} E.~L.,  2005, \mn@doi [Nature]
  {10.1038/nature03202}, \href
  {http://adsabs.harvard.edu/abs/2005Natur.433...45M} {433, 45}

\bibitem[\protect\citeauthoryear{{McNamara} et~al.,}{{McNamara}
  et~al.}{2006}]{McNamaraetal06}
{McNamara} B.~R.,  et~al., 2006, \mn@doi [ApJ] {10.1086/505859}, \href
  {https://ui.adsabs.harvard.edu/abs/2006ApJ...648..164M} {648, 164}

\bibitem[\protect\citeauthoryear{{Monteiro-Oliveira}, {Cypriano}, {Machado},
  {Lima Neto}, {Ribeiro}, {Sodr{\'e}}  \& {Dupke}}{{Monteiro-Oliveira}
  et~al.}{2017}]{Monteiro-Oliveiraetal17}
{Monteiro-Oliveira} R.,  {Cypriano} E.~S.,  {Machado} R.~E.~G.,  {Lima Neto}
  G.~B.,  {Ribeiro} A.~L.~B.,  {Sodr{\'e}} L.,   {Dupke} R.,  2017, \mn@doi
  [MNRAS] {10.1093/mnras/stw3238}, \href
  {https://ui.adsabs.harvard.edu/abs/2017MNRAS.466.2614M} {466, 2614}

\bibitem[\protect\citeauthoryear{{Mulroy} et~al.,}{{Mulroy}
  et~al.}{2019}]{Mulroyetal19}
{Mulroy} S.~L.,  et~al., 2019, \mn@doi [MNRAS] {10.1093/mnras/sty3484}, \href
  {https://ui.adsabs.harvard.edu/abs/2019MNRAS.484...60M} {484, 60}

\bibitem[\protect\citeauthoryear{{O'Sullivan}, {Giacintucci}, {David}, {Gitti},
  {Vrtilek}, {Raychaudhury}  \& {Ponman}}{{O'Sullivan}
  et~al.}{2011}]{OSullivanetal11b}
{O'Sullivan} E.,  {Giacintucci} S.,  {David} L.~P.,  {Gitti} M.,  {Vrtilek}
  J.~M.,  {Raychaudhury} S.,   {Ponman} T.~J.,  2011, \mn@doi [ApJ]
  {10.1088/0004-637X/735/1/11}, \href
  {http://adsabs.harvard.edu/abs/2011ApJ...735...11O} {735, 11}

\bibitem[\protect\citeauthoryear{{Okabe} \& {Smith}}{{Okabe} \&
  {Smith}}{2016}]{OkabeSmith16}
{Okabe} N.,  {Smith} G.~P.,  2016, \mn@doi [MNRAS] {10.1093/mnras/stw1539},
  \href {https://ui.adsabs.harvard.edu/abs/2016MNRAS.461.3794O} {461, 3794}

\bibitem[\protect\citeauthoryear{{Okabe}, {Takada}, {Umetsu}, {Futamase}  \&
  {Smith}}{{Okabe} et~al.}{2010}]{Okabeetal10}
{Okabe} N.,  {Takada} M.,  {Umetsu} K.,  {Futamase} T.,   {Smith} G.~P.,  2010,
  \mn@doi [PASJ] {10.1093/pasj/62.3.811}, \href
  {https://ui.adsabs.harvard.edu/abs/2010PASJ...62..811O} {62, 811}

\bibitem[\protect\citeauthoryear{{Owers}, {Nulsen}  \& {Couch}}{{Owers}
  et~al.}{2011}]{Owersetal11}
{Owers} M.~S.,  {Nulsen} P. E.~J.,   {Couch} W.~J.,  2011, \mn@doi [ApJ]
  {10.1088/0004-637X/741/2/122}, \href
  {https://ui.adsabs.harvard.edu/abs/2011ApJ...741..122O} {741, 122}

\bibitem[\protect\citeauthoryear{{Panagoulia}, {Fabian}, {Sanders}  \&
  {Hlavacek-Larrondo}}{{Panagoulia} et~al.}{2014}]{Panagouliaetal14b}
{Panagoulia} E.~K.,  {Fabian} A.~C.,  {Sanders} J.~S.,   {Hlavacek-Larrondo}
  J.,  2014, \mn@doi [MNRAS] {10.1093/mnras/stu1499}, \href
  {http://adsabs.harvard.edu/abs/2014MNRAS.444.1236P} {444, 1236}

\bibitem[\protect\citeauthoryear{{Postman} et~al.,}{{Postman}
  et~al.}{2012}]{Postmanetal12}
{Postman} M.,  et~al., 2012, \mn@doi [ApJS] {10.1088/0067-0049/199/2/25}, \href
  {https://ui.adsabs.harvard.edu/abs/2012ApJS..199...25P} {199, 25}

\bibitem[\protect\citeauthoryear{{Prasad} \& {Chengalur}}{{Prasad} \&
  {Chengalur}}{2012}]{PrasadChengalur12}
{Prasad} J.,  {Chengalur} J.,  2012, \mn@doi [Experimental Astronomy]
  {10.1007/s10686-011-9279-5}, \href
  {https://ui.adsabs.harvard.edu/abs/2012ExA....33..157P} {33, 157}

\bibitem[\protect\citeauthoryear{{Rafferty}, {McNamara}, {Nulsen}  \&
  {Wise}}{{Rafferty} et~al.}{2006}]{Raffertyetal06}
{Rafferty} D.~A.,  {McNamara} B.~R.,  {Nulsen} P.~E.~J.,   {Wise} M.~W.,  2006,
  \mn@doi [ApJ] {10.1086/507672}, \href
  {http://adsabs.harvard.edu/abs/2006ApJ...652..216R} {652, 216}

\bibitem[\protect\citeauthoryear{{Rieke} et~al.,}{{Rieke}
  et~al.}{2004}]{Riekeetal04}
{Rieke} G.~H.,  et~al., 2004, \mn@doi [ApJS] {10.1086/422717}, \href
  {https://ui.adsabs.harvard.edu/abs/2004ApJS..154...25R} {154, 25}

\bibitem[\protect\citeauthoryear{{Rieke}, {Alonso-Herrero}, {Weiner},
  {P{\'e}rez-Gonz{\'a}lez}, {Blaylock}, {Donley}  \& {Marcillac}}{{Rieke}
  et~al.}{2009}]{Riekeetal09}
{Rieke} G.~H.,  {Alonso-Herrero} A.,  {Weiner} B.~J.,  {P{\'e}rez-Gonz{\'a}lez}
  P.~G.,  {Blaylock} M.,  {Donley} J.~L.,   {Marcillac} D.,  2009, \mn@doi
  [ApJ] {10.1088/0004-637X/692/1/556}, \href
  {https://ui.adsabs.harvard.edu/abs/2009ApJ...692..556R} {692, 556}

\bibitem[\protect\citeauthoryear{{Rines}, {Geller}, {Diaferio}  \&
  {Kurtz}}{{Rines} et~al.}{2013}]{Rinesetal13}
{Rines} K.,  {Geller} M.~J.,  {Diaferio} A.,   {Kurtz} M.~J.,  2013, \mn@doi
  [ApJ] {10.1088/0004-637X/767/1/15}, \href
  {https://ui.adsabs.harvard.edu/abs/2013ApJ...767...15R} {767, 15}

\bibitem[\protect\citeauthoryear{{Roland}, {Hanisch}, {Veron}  \&
  {Fomalont}}{{Roland} et~al.}{1985}]{Rolandetal85}
{Roland} J.,  {Hanisch} R.~J.,  {Veron} P.,   {Fomalont} E.,  1985, A\&A, \href
  {https://ui.adsabs.harvard.edu/abs/1985A&A...148..323R} {148, 323}

\bibitem[\protect\citeauthoryear{{Sabater} et~al.,}{{Sabater}
  et~al.}{2019}]{Sabateretal19}
{Sabater} J.,  et~al., 2019, \mn@doi [A\&A] {10.1051/0004-6361/201833883},
  \href {https://ui.adsabs.harvard.edu/abs/2019A&A...622A..17S} {622, A17}

\bibitem[\protect\citeauthoryear{{Savini} et~al.,}{{Savini}
  et~al.}{2019}]{Savinietal19}
{Savini} F.,  et~al., 2019, \mn@doi [A\&A] {10.1051/0004-6361/201833882}, \href
  {https://ui.adsabs.harvard.edu/abs/2019A&A...622A..24S} {622, A24}

\bibitem[\protect\citeauthoryear{{Schellenberger}, {David}, {O'Sullivan},
  {Vtilek}  \& {Haines}}{{Schellenberger} et~al.}{2019}]{Schellenbergeretal19}
{Schellenberger} G.,  {David} L.,  {O'Sullivan} E.,  {Vtilek} J.~M.,   {Haines}
  C.~P.,  2019, \mn@doi [ApJ] {10.3847/1538-4357/ab35e4}, \href
  {https://ui.adsabs.harvard.edu/abs/2019ApJ...882...59S} {882, 59}

\bibitem[\protect\citeauthoryear{{Shin}, {Woo}  \& {Mulchaey}}{{Shin}
  et~al.}{2016}]{Shinetal16}
{Shin} J.,  {Woo} J.-H.,   {Mulchaey} J.~S.,  2016, \mn@doi [ApJS]
  {10.3847/1538-4365/227/2/31}, \href
  {https://ui.adsabs.harvard.edu/abs/2016ApJS..227...31S} {227, 31}

\bibitem[\protect\citeauthoryear{{Smol{\v{c}}i{\'c}}, {Finoguenov}, {Zamorani},
  {Schinnerer}, {Tanaka}, {Giodini}  \& {Scoville}}{{Smol{\v{c}}i{\'c}}
  et~al.}{2011}]{Smolcicetal11}
{Smol{\v{c}}i{\'c}} V.,  {Finoguenov} A.,  {Zamorani} G.,  {Schinnerer} E.,
  {Tanaka} M.,  {Giodini} S.,   {Scoville} N.,  2011, \mn@doi [MNRAS]
  {10.1111/j.1745-3933.2011.01092.x}, \href
  {https://ui.adsabs.harvard.edu/abs/2011MNRAS.416L..31S} {416, L31}

\bibitem[\protect\citeauthoryear{{Snios} et~al.,}{{Snios}
  et~al.}{2018}]{Sniosetal18}
{Snios} B.,  et~al., 2018, \mn@doi [ApJ] {10.3847/1538-4357/aaaf1a}, \href
  {https://ui.adsabs.harvard.edu/abs/2018ApJ...855...71S} {855, 71}

\bibitem[\protect\citeauthoryear{{Tremblay} et~al.,}{{Tremblay}
  et~al.}{2015}]{Tremblayetal15}
{Tremblay} G.~R.,  et~al., 2015, \mn@doi [MNRAS] {10.1093/mnras/stv1151}, \href
  {https://ui.adsabs.harvard.edu/abs/2015MNRAS.451.3768T} {451, 3768}

\bibitem[\protect\citeauthoryear{{Vrtilek}, {David}, {Grego}, {Jerius},
  {Jones}, {Forman}, {Donnelly}  \& {Ponman}}{{Vrtilek}
  et~al.}{2000}]{Vrtileketal00}
{Vrtilek} J.~M.,  {David} L.~P.,  {Grego} L.,  {Jerius} D.,  {Jones} C.,
  {Forman} W.,  {Donnelly} R.~H.,   {Ponman} T.~J.,  2000, in Constructing the
  Universe with Clusters of Galaxies.

\bibitem[\protect\citeauthoryear{{Wen}, {Han}  \& {Liu}}{{Wen}
  et~al.}{2012}]{Wenetal12}
{Wen} Z.~L.,  {Han} J.~L.,   {Liu} F.~S.,  2012, \mn@doi [ApJS]
  {10.1088/0067-0049/199/2/34}, \href
  {https://ui.adsabs.harvard.edu/abs/2012ApJS..199...34W} {199, 34}

\bibitem[\protect\citeauthoryear{{Wilson}, {Smith}  \& {Young}}{{Wilson}
  et~al.}{2006}]{Wilsonetal06}
{Wilson} A.~S.,  {Smith} D.~A.,   {Young} A.~J.,  2006, \mn@doi [ApJ]
  {10.1086/504108}, \href
  {https://ui.adsabs.harvard.edu/abs/2006ApJ...644L...9W} {644, L9}

\bibitem[\protect\citeauthoryear{{Wuyts} et~al.,}{{Wuyts}
  et~al.}{2011}]{Wuytsetal11}
{Wuyts} S.,  et~al., 2011, \mn@doi [ApJ] {10.1088/0004-637X/742/2/96}, \href
  {http://adsabs.harvard.edu/abs/2011ApJ...742...96W} {742, 96}

\bibitem[\protect\citeauthoryear{{York} et~al.,}{{York}
  et~al.}{2000}]{Yorketal00}
{York} D.~G.,  et~al., 2000, \mn@doi [AJ] {10.1086/301513}, \href
  {https://ui.adsabs.harvard.edu/abs/2000AJ....120.1579Y} {120, 1579}

\bibitem[\protect\citeauthoryear{{Yu} et~al.,}{{Yu} et~al.}{2018}]{Yuetal18}
{Yu} H.,  et~al., 2018, \mn@doi [ApJ] {10.3847/1538-4357/aaa421}, \href
  {https://ui.adsabs.harvard.edu/abs/2018ApJ...853..100Y} {853, 100}

\bibitem[\protect\citeauthoryear{{Yun}, {Reddy}  \& {Condon}}{{Yun}
  et~al.}{2001}]{Yunetal01}
{Yun} M.~S.,  {Reddy} N.~A.,   {Condon} J.~J.,  2001, \mn@doi [ApJ]
  {10.1086/323145}, \href
  {https://ui.adsabs.harvard.edu/abs/2001ApJ...554..803Y} {554, 803}

\bibitem[\protect\citeauthoryear{{ZuHone} \& {Sims}}{{ZuHone} \&
  {Sims}}{2019}]{ZuHoneSims19}
{ZuHone} J.~A.,  {Sims} J.,  2019, \mn@doi [ApJ] {10.3847/1538-4357/ab2b34},
  \href {https://ui.adsabs.harvard.edu/abs/2019ApJ...880..145Z} {880, 145}

\makeatother
\end{thebibliography}
